\documentclass[aps,twocolumn,showpacs,floatfix]{revtex4}

\pdfoutput=1

\usepackage{amsmath,amssymb,bm}
\usepackage{graphicx}
\usepackage{latexsym}

\def\be{\begin{equation}}
\def\ee{\end{equation}}
\def\bea{\begin{eqnarray}}
\def\eea{\end{eqnarray}}
\def\ba{\begin{array}}
\def\ea{\end{array}}

\begin{document}

\title{Glassiness in Uniformly Frustrated Systems}
\author{Maxim Dzero}
\affiliation{Condensed Matter Theory Center and Department of Physics, University of
Maryland, College Park, MD 20742}
\affiliation{Department of Physics, Kent State University, Kent, OH 44242}
\author{ J\"{o}rg Schmalian}
\affiliation{Department of Physics and Astronomy and Ames Laboratory, Iowa State
University, Ames, IA 50011}
\author{Peter G. Wolynes}
\affiliation{Department of Chemistry and Biochemistry and Department of Physics,
University of California, San Diego, La Jolla, CA 92093 }
\date{\today }

\begin{abstract}
We review several models of glassy systems where the randomness is self
generated, i.e. already an infinitesimal amount of disorder is sufficient to
cause a transition to a non-ergodic, glassy state. We discuss the
application of the replica formalism developed for the spin glass systems to
study the glass transition in uniformly frustrated many-body systems. Here a
localization in configuration space emerges leading to an entropy crisis of
the system. Using a combination of density functional theory and Landau
theory of the glassy state, we first analyze the mean field glass transition
within the saddle point approximation. We go beyond the saddle point
approximation by considering the energy fluctuations around the saddle point
and evaluate the barrier height distribution.
\end{abstract}

\pacs{75.10.Nr, 05.70.Ln}
\maketitle

\section{Introduction}

The glass transition and slow glassy dynamics are phenomena most
widely studied in the context of supercooled liquids and polymer melts
\cite%
{Angel96,ang88}. Important progress has been made in understanding the
complexity of these phenomena by combining dynamical approaches with
the concept of an underlying energy landscape, as is discussed in
detail in other chapters of this volume. For example, on the level of
the mean field theory of glasses, it has been demonstrated that the
dynamical, ideal mode coupling theory \cite{mc} and energy landscape
based replica mean field theories describe the same underlying
physics, yet from rather different perspectives
\cite{KT87,KW87}. Novel replica approaches have been developed to
characterize the emergence of a metastable amorphous solid \cite%
{Mon95,Franz95,MePa98.1,MePa98.2}. In addition, droplet arguments have
been proposed to include important physics beyond the pure mean field
description and led to the formulation of the random first order
transition theory of glasses \cite{KTW89}. The latter is particularly
important if one wants to make specific predictions for experiments
that require one to go beyond the mean field limit
\cite{XW00,XW01,Lubchenko01,Lubchenko04a,Biroli04}. This random first
order transition theory of glasses with an underlying entropy crisis
\cite{Kauzmann48} offers a general and quantitative description of
structural glasses. In addition, it defines a universality class for
complex many body systems that is of importance for a much larger
class of materials. 
\begin{figure}[h]
\includegraphics[scale=0.28,angle=0]{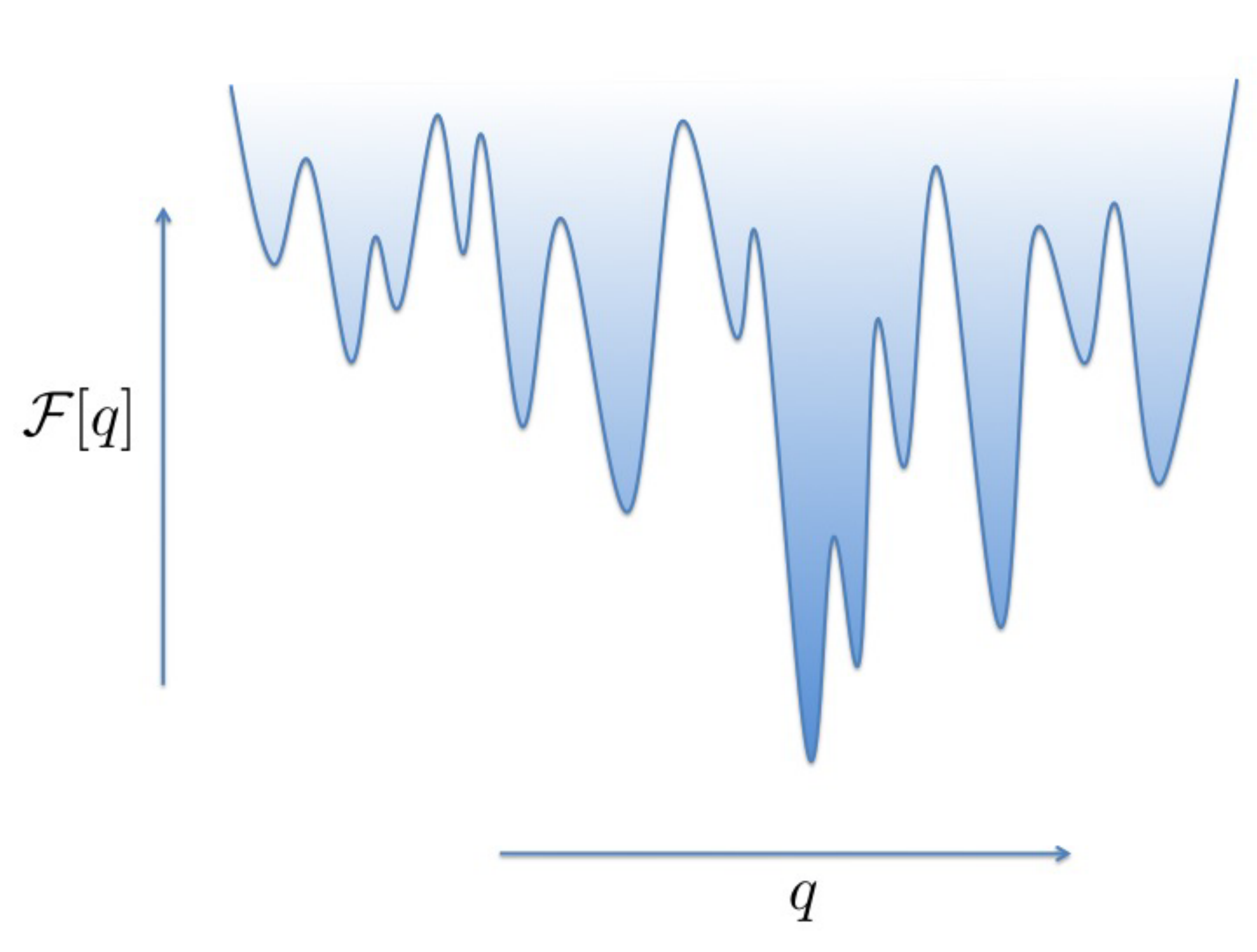}
\caption{Free energy of a system with entropy crisis as a function of the
generalized coordinate $q$. For a specific case of a structural glass $q$
plays a role of density of the local atomic configurations. Glassy behavior
is characterized by the emergence of large number of metastable states
separated by high energy barriers.}
\label{Fig1}
\end{figure}

\begin{figure}[h]
\includegraphics[scale=0.28,angle=0]{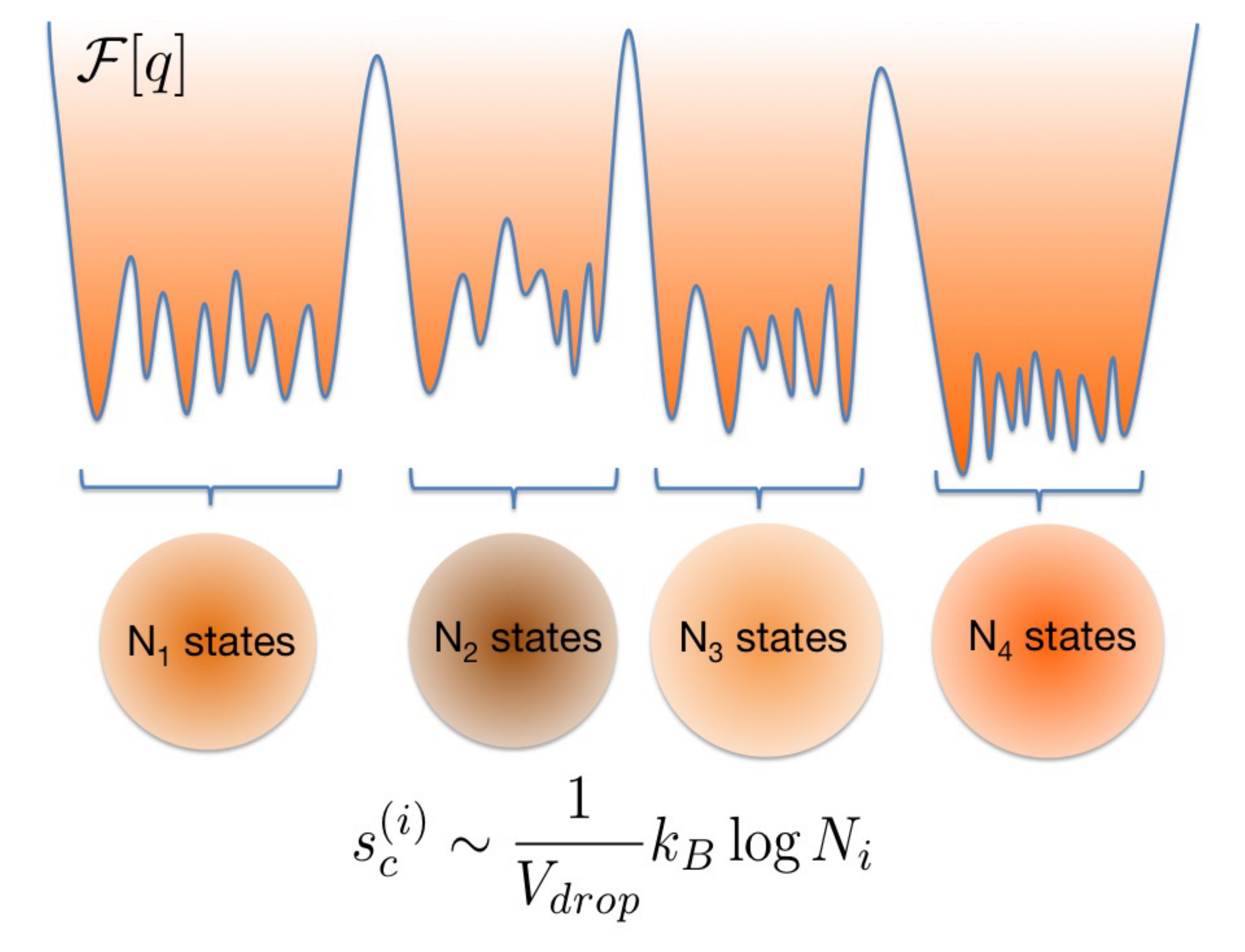}
\caption{Emergence of an exponentially large number of metastable
configurations and the possibility for a system to realize these
configurations gives rise to the configurational entropy. The latter serves
as the driving force for the structural transitions in the glassy phase.}
\label{Fig2}
\end{figure}

Glassy dynamics usually occurs when a system is supercooled below a first
order transition to an ordered state. In supercooled liquids this ordered
state is the crystalline solid. If one applies the same ideas to systems
with charge or spin collective degrees of freedom in a correlated material,
the ordered state corresponds to an electron crystal or an ordered magnetic
state. The equilibrium state is therefore ordered. Supercooling the liquid
state is possible once the nucleation barriers are high. Thus, in what
follows we ignore this ordered state and consider time scales that are short
compared to the nucleation time $t_{\mathrm{nucl}}$. One can now ask the
question whether and for how long one can still consider the supercooled
liquid in local equilibrium. Here, by local equilibrium we mean that the
system is ergodic and all configurations are being explored with a
probability given by the Boltzmann distribution (only excluding the sharp
peak that corresponds to the ordered state). One expects that this
supercooled state will fall out of equilibrium over very long time scales,
i.e. behave nonergodically, once the barriers between distinct
configurations of comparable energy become very large, Fig. \ref{Fig1}. If
this happens, the system will explore only a subset of the available phase
space and will eventually freeze. Suppose there are $\mathcal{N}_{ms}$ of
such metastable configurations, that are separated by high barriers, Fig. %
\ref{Fig2}. A calorimetric measurement will then yield an entropy that is
reduced by 
\begin{equation}
S_{c}\simeq k_{B}\log \mathcal{N}_{ms}
\end{equation}%
compared to the ergodic situation, where $S_{c}$ is referred to as the
configurational entropy. If $\mathcal{N}_{ms}$ is exponentially large in the
system size, $S_{c}$ becomes extensive and the dynamically frozen state
strictly speaking does not obey the third law of thermodynamics. For liquids
it appears that $S_c$ on infinite time scales would extrapolate to zero at a
temperature $T_K>0$. Extrapolating further to $T<T_K$ would give $S_c$ less
then zero. This is the entropy crisis of Kauzmann [16]. Such a crisis is
avoided by an ideal glass transition in the models highlighted here.

A crucial question to explore is under what circumstances the
configurational entropy is a well defined and meaningful
concept. Metastable states become sharp conceptually only within a
mean field theory description, where barriers can be infinitely high
and a system might get trapped in a local minimum of phase space
forever. Within mean field theory, the number of states that are
separated by high barriers proliferates at the mode coupling
temperature $T_{A}$. The sudden onset of $S_{c}\left( T\right) $ at
$T=T_{A}$ should not be misunderstood to mean that new configurations
emerge at that temperature. These configurations were present already
in the liquid state above $T_{A}$. What happens at $T_{A}$ is that the
barriers grow and that the system's dynamics becomes \emph{localized
  in configuration space}.  It is this localization in configuration
space that is the key element of the random first order theory (not
the ultimate glass transition). Within mean field theory the system
freezes right at $T_{A}$, while activated events, that are a key
element of the theory, allow for a slowed down dynamics even below
$T_{A}$, reflecting the fact that those barriers are large but finite.

One might criticize this approach since going beyond mean field theory
does not allow for a sharp thermodynamic definition of
$S_{c}$. However, there are well known examples that demonstrate the
opposite: understanding overheated or supercooled phases close to a
first order transition also starts from the mean field concept of a
local minimum in the energy landscape. Once supplemented with an
appropriate droplet theory of the metastable state it yields a
description of the nucleation dynamics at first order
transitions\cite{Langer67}. The logic used in the random first order
theory of glasses is very similar. One first develops a mean field
theory and then supplements it by an appropriate droplet calculation
that clearly goes beyond strict mean field theory.

\begin{figure}[h]
\includegraphics[scale=0.28,angle=0]{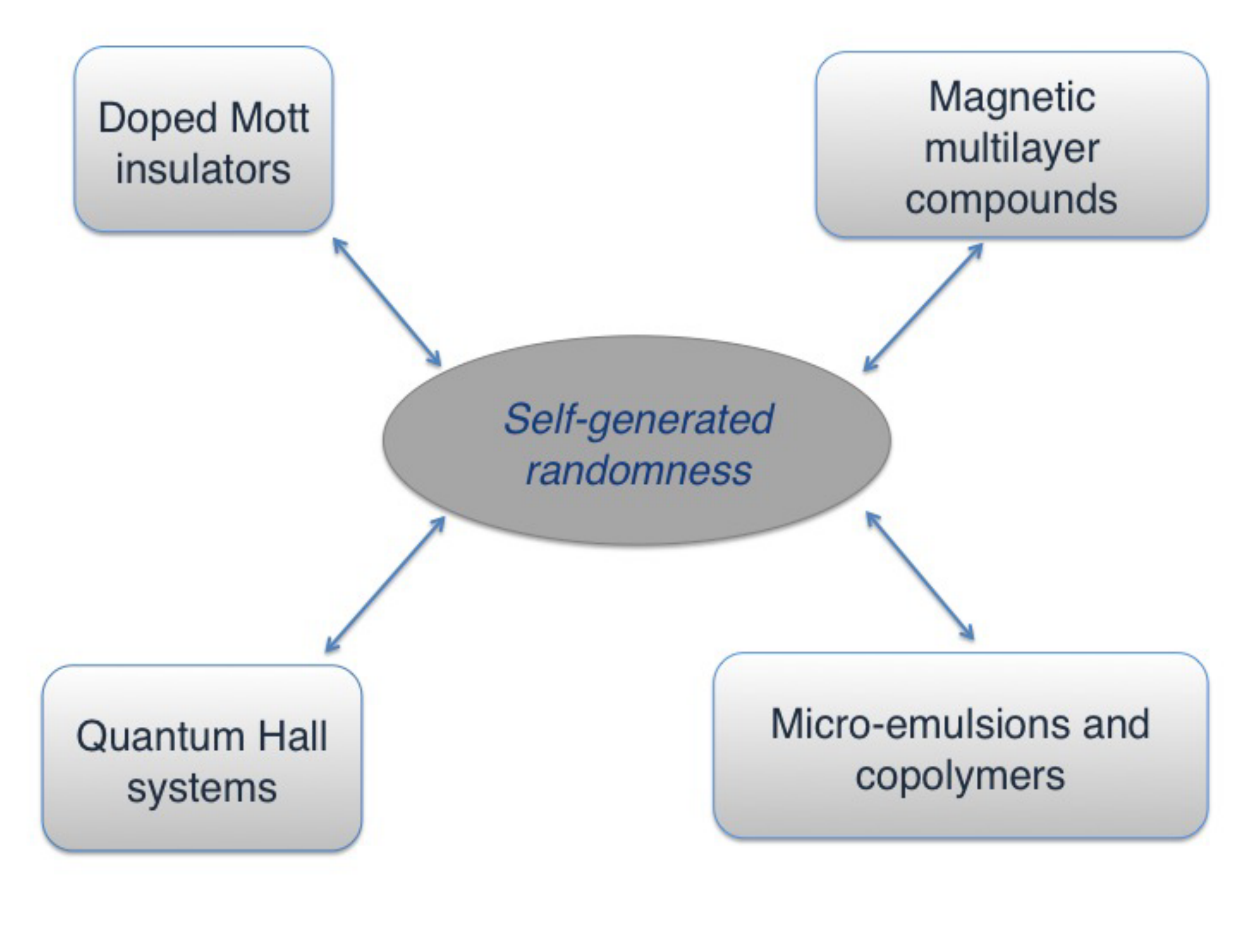}
\caption{Phenomenon of self-generated randomness appears in a variety of
physical systems ranging from strongly correlated two-dimensional electrons
to polymer melts and mesoscopic self-assembly structures.}
\label{Fig3}
\end{figure}

There are numerous indications that the underlying principles that govern
glassy behavior in supercooled liquids apply to other many body systems as
well. Generally, glassy systems are characterized by slow relaxations and a
broad spectrum of excitations, a behavior found in a number of strongly
correlated hard condensed matter physics systems. \ A particularly
interesting situation occurs when interactions, that change very differently
with distance, compete with one another. In such \emph{uniformly frustrated
systems} an intermediate length scale can emerge that leads to new spatial
structures and inhomogeneities. A classical example is a ferromagnet where
finite range exchange interactions compete with the long range dipole-dipole
interactions and leads to the formation of magnetic domains. Similar
behavior is in fact abundant in a number of correlated electron systems.
Examples are stripe formation in doped Mott insulators \cite%
{stripe,stripe2,mang01,Millis96,Dagotto,nmr01,nmr02,nmr03,nmr3b,nmr04,msr01,msr02}%
, defect formation in two-dimensional electron systems \cite%
{Popovich07,Spivak2006,2DSpivak2010}, bubbles of electronic states of high
Landau levels in quantum Hall systems \cite{QHE,QHE2}, magnetic domains in
magnetic multilayer compounds \cite{magmtlyr}, and mesoscopic structures
formed in self-assembly systems \cite{selfassbl}. These systems typically
exhibit a multi-time-scale dynamics similar to the relaxation found in
glasses. This observation suggests that glassy behavior and large relaxation
times are caused by the competition of interactions with different
characteristic length scales \cite%
{Popovic2002,Popovic2,Vlad2002,Denis2002,Ohta86,Kivelson,Nussinov,Schmalian,SWW00,WSW03,WSW04}%
, and not primarily due to the presence of strong disorder in the system. As
we will see, this does not mean that disorder is irrelevant for uniformly
frustrated systems. In fact, the opposite is true. We find that even the
smallest amounts of disorder and imperfections drive such system into a
non-ergodic regime that otherwise emerges only in the presence of very
strong disorder. In the language of a renormalization group flow this means
that the strength of disorder, $g$, flows to larger and larger values \ and
reaches a strong-disorder fixed point $g^{\ast }$, even if the physical,
bare disorder strength is very small. Glassiness arises spontaneously even
for infinitesimal extrinsic disorder, leading to \textit{self-generated
randomness}, Fig. \ref{Fig3}. In uniformly frustrated systems self generated
randomness and the emergence of an entropy crisis were first discussed in
Ref. \cite{SWW00}. For a beautiful example of self generated randomness in
the context of coupled Josephson junction arrays, see Ref. \cite{CIS98}. 

While frustration through competing interactions is easiest to analyze, the
notion that uniform geometric frustration is at the heart of the structural
glass transition was already discussed and analyzed in Refs.\cite%
{Sadoc,Nelson,Sethna} with the underlying view that one can capture
frustration through an underlying ideal structure in curved space as a
reference state. Recently this general idea was analyzed in the specific
context of glass formation in a hyperbolic space\cite{Sausset}. Modeling
frustrated interactions by a competing interactions on different length
scales was suggested in Ref.\cite{DKivelson} where the concept of avoided
criticality for uniformly frustrated systems was introduced, see also Refs.
\cite{Chayes,Nussinov} for further details. Evidence for fragile
glass-forming behavior in the relaxation of Coulomb frustrated
three-dimensional systems by means of a Monte Carlo simulation was given in
Ref.\cite{Grousson}. Furthermore, in Ref.\cite{GroussonMC} the close
relation between the replica approach presented here and the results
obtained within a mode coupling analysis of a uniformly frustrated system,
was demonstrated explicitly. For a recent review on the subject, see Ref. \cite{Tarjus}

In this chapter we discuss in some detail model systems that display self
generated randomness. We do so by applying the replica formalism originally
developed to describe disordered spin glasses \cite{Mon95} and structural
glasses \cite{MePa98.1,MePa98.2}. We demonstrate that there are many body
systems that undergo a mean field glass transition for arbitrarily weak
disorder \cite{Schmalian,SWW00,WSW03,WSW04}. We then develop a Landau theory
of the glassy state that reproduces the key physical behavior of these
uniformly frustrated systems \cite{Dzero05,Dzero09}. The appeal of the
Landau theory is that it easily permits for a generalization beyond the mean
field limit, where we include instanton events to describe dynamical
heterogeneity in glassy systems.

\section{Uniformly frustrated systems: a model Hamiltonian approach to glass
formation}

We summarize the main featured of a simple model that exhibits glassy
behavior due to infinitesimal amount of disorder \cite%
{Schmalian,SWW00,WSW03,WSW04}. The model exhibits competition of
interactions on different length scales and there are no explicitly quenched
degrees of freedom. Consider $\varphi (\mathbf{x)}$: the spatially varying
amplitude that characterizes the collective degrees of freedom of a many
body system. Depending on the problem under consideration $\varphi \left( 
\mathbf{x}\right) $ corresponds to the magnetization $m\left( \mathbf{x}%
\right) $ (in case of magnetic domain formations) or a density fluctuation
relative to the mean density, $\rho \left( \mathbf{x}\right) =\rho
_{0}+\varphi \left( \mathbf{x}\right) $. Here, $\rho \left( \mathbf{x}%
\right) $ stands for the electron density in case of a doped Mott insulator 
\cite{Kivelson,Nussinov,Schmalian} or for the density of amphiphilic
molecules in case of a microemulsion or a copolymer \cite%
{Leibler,Fredrickson,Deem94,WWSW02,Zhang06,Wu09}.

We consider the effective Hamiltonian%
\begin{equation}
\mathcal{H}=\mathcal{H}_{0}+\mathcal{H}_{\mathrm{int}}  \label{H}
\end{equation}%
where $\mathcal{H}_{0}$ contains all terms of second order in $\varphi (%
\mathbf{x)}$. Specifically, we consider 
\begin{eqnarray}
\mathcal{H}_{0} &=&\frac{1}{2}\int d^{d}x\left( r_{0}\varphi^2 (\mathbf{x)} +%
\left[\nabla \varphi (\mathbf{x)}\right] ^{2}\right)  \notag \\
&&+\frac{1}{2}\int d^{d}x\int d^{d}x^{\prime }\varphi (\mathbf{x)}V\left(
\left\vert \mathbf{x-x}^{\prime }\right\vert \right) \varphi (\mathbf{x}%
^{\prime }).  \label{H0}
\end{eqnarray}%
For the non-linear interaction term we assume the local form 
\begin{equation}
\mathcal{H}_{\mathrm{int}}=\int d^{d}xW\left( \varphi \left( \mathbf{x}%
\right) \right)  \label{Hint}
\end{equation}%
where we use\ 
\begin{equation}
W\left( \varphi \right) =\frac{u_{3}}{3}\varphi ^{3}+\frac{u_{4}}{4}\varphi
^{4}.  \label{exp}
\end{equation}

In Eq. (\ref{H0}), $r_{0}<0$ favors local order of $\varphi
(\mathbf{x)}$.  Thus, in the absence of the long distance coupling
$V\left( \mathbf{x-x}%
  ^{\prime }\right) $ the system is expected to be homogeneously
ordered. The gradient term favors homogeneous configurations of
$\varphi \left( \mathbf{x}%
\right) $. It is written by taking the the continuum limit \ of an
underlying short range interaction. \ In the case where $\varphi
\left( \mathbf{x%
  }\right) $ refers to the deviation from the mean density, the
average $%
\left\langle \varphi \left( \mathbf{x}\right) \right\rangle $ vanishes
by construction. Still the expected low temperature minimum of the
free energy corresponds to configurations where regions of macroscopic
size have constant field values ($\varphi \left( \mathbf{x}\right)
\simeq const.$), separated by domain walls of minimal area, due to the
presence of the gradient term. This is in fact a toy model for
macroscopic phase separation.  The corresponding ordering temperature
can be estimated within mean field theory. For $d=3$ it follows, for
example, $T_{c}^{0}=\frac{2\pi ^{2}\left\vert r_{0}\right\vert
}{u_{4}\Lambda }$, with momentum cutoff $%
\Lambda $ of the order of an inverse lattice constant.

The situation changes significantly once we include an additional long range
interaction 
\begin{equation}
V\left( x\right) =l_{0}^{-\left( d+2\right) }\left( \frac{l_{0}}{x}\right)
^{\tau }.
\end{equation}%
Here $l_{0}$ is the new typical length scale that results from the
competition of the gradient term and the long range interaction with $0<\tau
<d$. \ The long range interaction vanishes in the limit $l_{0}\rightarrow
\infty $. The exponent $\tau $ determines the rate of decay of the
interaction. The effect of this long range interaction can easily be seen in
momentum space. The Fourier transform of $V\left( x\right) $ is given as 
\begin{equation}
V\left( k\right) =C_{d}\frac{l_{0}^{-2}}{\left( l_{0}k\right) ^{d-\tau }},
\label{mom}
\end{equation}%
with dimensionless coefficient $C_{d}=2^{d-\tau }\pi ^{d/2}\Gamma \left( 
\frac{d-\tau }{2}\right) /\Gamma \left( \tau /2\right) $. For $\tau <d$,
homogeneous field configurations with $k\rightarrow 0$ are energetically
very costly and spatial pattern with finite wave number $k_{0}\sim 2\pi
/l_{0}$ emerge as consequence of the competition between the short range
interaction and long range forces. For $\tau =d$ , $V\left( k\right) \propto
\log \left( kl_{0}\right) $ and homogeneous configurations are still
suppressed, albeit only logarithmically. This is relevant for the
dipole-dipole interactions in $d=3$.

One can also consider situations with a screened interaction, such as 
\begin{equation}
V_{scr}\left( x\right) =l_{0}^{-\left( d+2\right) }\left( \frac{l_{0}}{x}%
\right) ^{\tau }\exp \left( -x/a_{scr}\right) .
\end{equation}%
As shown in Ref. \cite{WWSW02}, glass formation is virtually unchanged
compared to the unscreened interaction once $l_{0}\ll a_{scr}$, while it is
suppressed in the opposite limit. The theory of Refs. \cite{SWW00,WWSW02}
was also applied to study the glassy phase in a model for charged colloids 
\cite{Tarzia06,Tarzia07}.

At the Hartree level, the correlation function that results from these two
competing interactions is 
\begin{equation}
G_{H}(k)=\frac{1}{k^{2}+r^{\prime }+V\left( k\right) },
\end{equation}%
with $r^{\prime }=r_{0}+u_{4}T\int \frac{d^{d}q}{(2\pi )^{d}}\,G_{H}\left(
k\right) $. The intermediate length scale that results from the competition
between short and long range interactions leads to a peak in the correlation
function at 
\begin{equation}
k_{0}=\widetilde{C}_{d}l_{0}^{-1}
\end{equation}%
where $\widetilde{C}_{d}=\left( \frac{C_{d}\left( d-\tau \right) }{2}\right)
^{\frac{1}{2+d-\tau }}$ . We can approximate $G_{H}(k)$ close to $k_{0}$ as: 
\begin{equation}
G_{H}(k)=\frac{A_{0}}{\left( k-k_{0}\right) ^{2}+\xi ^{-2}}  \label{Braz}
\end{equation}%
where $A_{0}=\left( 2+d-\tau \right) ^{-1}$. The liquid state correlation
length is given by $\xi ^{-2}=A_{0}r^{\prime }+k_{0}^{2}$. Its temperature
dependence must be determined from a conventional equilibrium calculation.
The form Eq. (\ref{Braz}) holds for a large class of models with competing
interactions. For $u_{3}=0$ Brazovskii showed that such a model undergoes a
fluctuation induced first order transition to a state with lamellar, or
stripe order \cite{Brazovskii}. In general, this model has been discussed in
the context of complex crystallization \cite{Alexander78,BDM87}. For
example, Alexander and McTague \cite{Alexander78} argued that
crystallization of body centered cubic crystals is preferred if the
first-order character is not too pronounced. As shown in Refs. \cite%
{Groh99,Klein01} the preference for bcc order is rather for metastable
states that form near the spinodal. Below we will ignore ordered crystalline
states and assume that their nucleation kinetics is sufficiently slow to
supercool the disordered state.

Finally, we mention that our model, Eqs. (\ref{H}-\ref{Hint}), can also be
used to describe interacting liquids consisting of $N$-components \cite%
{WSW04}. In this case $\varphi =\left( \varphi _{1},\cdots ,\varphi
_{N}\right) $ is a vector whose components refer to the density deviations
of the different components of the liquid, with mean density $\rho _{0,l}$.
Now, the Gaussian part 
\begin{equation}
\begin{split}
\mathcal{H}_{0}=&\frac{1}{2}\int d^{d}xd^{d}x^{\prime } \\
\times& \sum_{ll^{\prime }}\varphi _{l}\left( \mathbf{x}\right) \left[
c_{ll^{\prime }}\left( \left\vert \mathbf{x}-\mathbf{x}^{\prime }\right\vert
\right) +\delta _{ll^{\prime }}\rho _{0,l}^{-1}\right] \varphi _{l^{\prime
}}\left( \mathbf{x}^{\prime }\right)
\end{split}%
\end{equation}%
is determined by the direct correlation function $c_{ll^{\prime }}\left(
x\right) $ of the fluid \cite{Hansen}. Within the density functional
approach pioneered by Ramakrishnan and Youssoff \cite{RY79} the nonlinear
part of the Hamiltonian is determined by the ideal gas free energy: 
\begin{equation}
W\left( \varphi \right) =\sum_{l}\left[ \left( \rho _{0,l}+\varphi
_{l}\right) \left[ \log \left( \frac{\rho _{0,l}+\varphi _{l}}{\rho _{0,l}}%
\right) -1\right] -\frac{\varphi _{l}{}^{2}}{2\rho _{0,l}}\right]
\end{equation}%
In case of the density functional theory, Eq. (\ref{exp}) corresponds
to the resummed virial expansion of the liquid.

In summary, models of the form Eqs. (\ref{H}-\ref{Hint}) capture the physics
of a large class of systems where competing interactions on different length
scales lead to complex spatial pattern and potentially to slow glassy
dynamics. In the next section we discuss in some detail the methods that
will be used to describe and analyze glassy systems. Those methods will then
be applied to the model of Eqs. (\ref{H}-\ref{Hint}) further below.

\section{ Entropy crisis and mean field formalism}

We first outline the main idea of the generalized replica mean field
approach\ for systems with an entropy crisis \cite{Mon95,MePa98.1}. We start
from a classical field theory with Hamiltonian, $H[\varphi ]$, and field
variable, $\varphi \left( \mathbf{x}\right) $, that yields the partition
function 
\begin{equation}
Z=\int D\varphi \exp \left( -\beta \mathcal{H}[\varphi ]\right) .
\label{psum_class}
\end{equation}%
The thermodynamic free energy is then given as $F=-T\log Z$. \ Suppose we
know the thermodynamic density of states 
\begin{equation}
\rho \left( f\right) =\int D\varphi \delta \left( f-\frac{1}{V}\mathcal{H}%
[\varphi ]\right) ,
\end{equation}%
where $V$ is the system volume. It follows with $s_{c}(f)=\frac{1}{V}\log
\rho \left( f\right) $ that 
\begin{equation}
Z=\int_{f_{\min }}^{f_{\max }}df\exp \left[ -V\left( \beta
f-s_{c}(f,T)\right) \right] .
\end{equation}%
For large $V$ the integral is evaluated at the saddle point, leading to the
free energy density 
\begin{equation}
F=-T\log Z\approx V\min_{f}\left[ f-Ts_{c}(f)\right] .
\end{equation}%
In the case when the number of distinct configurations with given
energy $f$ is exponentially large, $s_{c}\left( f\right) $ will be
finite as $V\rightarrow \infty $. \ Due to the configurational entropy
density, $s_{c}$, the free energy density then differs from its
minimum value, $f_{\min }$, even at the saddle point level. In mean
field theory $s_{c}\left( T\right) $ is finite below a temperature
$T_{A}$ that coincides with the dynamic mode coupling temperature
\cite{KT87}. \ At a lower temperature, $T_{K}<T_{A}$, the entropy
density vanishes like
\begin{equation}
s_{c}(T)\propto T-T_{K}+\mathcal{O}\left( \left( T-T_{K}\right) ^{2}\right)
\label{confent}
\end{equation}%
and the system undergoes a transition into a statically frozen state.

In order to have an explicit expression for $s_{c}$, it is convenient to
introduce the partition function\cite{MePa98.1,MePa98.2} 
\begin{equation}
Z(m)=\int_{f_{\min }}^{f_{\max }}df\exp \left[ -V\left( \beta
mf-s_{c}(f,T)\right) \right] ,
\end{equation}%
leading to the free energy density

\begin{equation}
f(m)=-\frac{T}{mV}\log Z(m)\approx \min_{f}\left[ f-Ts_{c}(f,T)/m\right] ,
\label{m}
\end{equation}%
such that $\overline{f}=\left. \frac{\partial mf(m)}{\partial m}\right\vert
_{m=1}$ gives the most probable value of $f$\ which might also be written as 
\begin{equation}
\overline{f}=\frac{1}{Z}\int_{f_{\min }}^{f_{\max }}df\text{ }f\text{ }\exp %
\left[ -V\left( \beta f-s_{c}(f,T)\right) \right] .  \label{ftil0}
\end{equation}%
Furthermore, the configurational entropy density \ follows from 
\begin{equation}
s_{c}=\left. \frac{1}{T}\frac{\partial f(m)}{\partial m}\right\vert _{m=1}.
\label{conf1}
\end{equation}
Next we will describe the ways of calculating the configurational entropy
density $s_c$.

\subsection{Replica formalism}

A systematic approach to calculate $F(m)=Vf\left( m\right) $ was developed
in Ref. \cite{Mon95}. \ The procedure of Ref. \cite{Mon95} also reproduces
the correct results for the configurational entropy in several random spin
systems. Furthermore, it allows for a rather transparent motivation for the
introduction of the variable $m$ of Eq. (\ref{m}). In what follows, we
summarize the main idea and some technical steps of this formalism.

We consider the partition function in the presence of a bias configuration $%
\psi \left( \mathbf{x}\right) $:%
\begin{equation}
Z\left[ \psi \right] =\int D\varphi e^{-\beta \mathcal{H}\left[ \varphi %
\right] -g\int d^{d}x\left( \varphi \left( \mathbf{x}\right) -\psi \left( 
\mathbf{x}\right) \right) ^{2}}.  \label{Zpsi}
\end{equation}%
Here, $\int D\varphi ...$ corresponds to the statistical sum over all
density configurations of the system. At the end of our calculation we
will take the limit $g\rightarrow 0$, however only after we take the
thermodynamic limit. For any finite $g$, the configuration $\psi
\left( \mathbf{x}\right) $ enters the problem as a quenched degree of
freedom, analog to random fields in disordered spin systems. We note
that in distinction to the usual random field problem, the probability
distribution function of metastable configurations $P\left[ \psi
\right]$ is not expressed in terms of uncorrelated random numbers, but
instead is determined by the partition function $Z\left[ \psi \right]
$ (see below).

The free energy for a given bias configuration is 
\begin{equation}
F\left[ \psi \right] =-T\log Z\left[ \psi \right] .
\end{equation}%
Physically $F\left[ \psi \right] $ can be interpreted as the free energy for
a metastable amorphous field configuration $\psi \left( \mathbf{x}\right) $. 
$Z\left[ \psi \right] $ in Eq. (\ref{Zpsi}) is obviously dominated by
configurations $\varphi \left( \mathbf{x}\right) \simeq \psi \left( \mathbf{x%
}\right) $ that correspond to local minima of $\mathcal{H}\left[ \varphi %
\right] $. In the replica formalism, no specific amorphous configuration
need be specified in order to perform the calculation. Rather, the
assumption is made that the probability distribution for metastable
configurations is determined by $F\left[ \psi \right] $ according 
\begin{equation}  \label{distrms}
P\left[ \psi \right] \propto \exp \left( -\beta _{\mathrm{eff}}F\left[ \psi %
\right] \right)
\end{equation}%
and is characterized by the effective temperature
$k_{B}T_{\mathrm{eff}%
}=\beta _{\mathrm{eff}}^{-1}\geq k_{B}T$. \ This allows one to
determine the mean free energy
\begin{equation}
\overline{F}=\int D\psi F\left[ \psi \right] P\left[ \psi \right]
\end{equation}%
and the corresponding mean configurational entropy 
\begin{equation}
S_{c}=-\int D\psi P\left[ \psi \right] \log P\left[ \psi \right] .
\label{Sc}
\end{equation}

It is physically appealing then to introduce the free energy difference, $%
\delta F$, via 
\begin{equation}
F=\overline{F}-\delta F,
\end{equation}%
where $\delta F$ gives the amount of energy lost if the system is trapped
into locally stable states and hence not able to explore the entire phase
space of the ideal thermodynamic equilibrium. If the limit $g\rightarrow 0$
behaves perturbatively, $\delta F=0$. This indicates that the number of
locally stable configurations stays finite in the thermodynamic limit, or at
least grows less rapid than exponential with $V$. In this case all states
are kinetically accessible. On the other hand, if the limit $g\rightarrow 0$
does not behave perturbatively, it means that the number of locally stable
states,\ $\mathcal{N}_{\mathrm{ms}}$,$\ $\ is exponentially large in $V$.
This allows us to identify the difference between the equilibrium and
typical free energy as an entropy: 
\begin{equation}
\delta F=TS_{c}.  \label{class}
\end{equation}%
The configurational entropy, $S_{c}=\log \mathcal{N}_{\mathrm{ms}}$,
is extensive if there are exponentially many metastable states. Within
mean field theory, where the barriers are infinitely high, \ the
emergence of $S_c$ renders the system incapable of exploring the
entire phase space. $S_{c}$ is then the amount of entropy which the
system that freezes it into a glassy state appears to lose due to its
nonequilibrium-dynamics.

The approach of Ref. \cite{Mon95} was successfully used to develop a mean
field theory for\ glass formation in supercooled liquids \cite%
{MePa98.1,MePa98.2} yielding results in detailed agreement with earlier,
non-replica approaches \cite{Stoessel84,Singh85}. \ The mean values $%
\overline{F}$ and $\overline{S_{c}}$ can be determined from a replicated
partition function 
\begin{equation}
Z\left[ m\right] =\int D\ \psi Z^m\left[ \psi \right]  \label{Zm}
\end{equation}%
via $\overline{F}=\frac{\partial }{\partial m}mF\left( m\right) $ and $S_{c}=%
\frac{m^{2}}{T}\frac{\partial }{\partial m}F\left( m\right) $ with 
\begin{equation}
F\left( m\right) =-\frac{T}{m}\log Z\left( m\right)
\end{equation}%
and replica index $m=\frac{T}{T_{\mathrm{eff}}}$. Inserting $Z\left[ \psi %
\right] $ of Eq. (\ref{Zpsi}) into Eq. (\ref{Zm}) and integrating over $\psi
,$ one gets\ 
\begin{equation}
Z\left[ m\right] =\int D^{m}\varphi e^{-\beta \sum_{a=1}^{m}\mathcal{H}\left[
\varphi _{a}\right] +g\sum_{a,b=1}^{m}\int d^{d}x\varphi _{a}\left( x\right)
\varphi _{b}\left( x\right) }  \label{Zmm}
\end{equation}%
which has a structure similar to a conventional equilibrium partition
function. The ergodicity breaking field $\psi $ causes a coupling between
replicas which might spontaneously lead to order in replica space even as $%
g\rightarrow 0$. This spontaneous coupling between different copies of the
system is then associated with \ a finite $S_{c}$ and thus glassiness.

\begin{figure}[h]
\includegraphics[scale=0.28,angle=0]{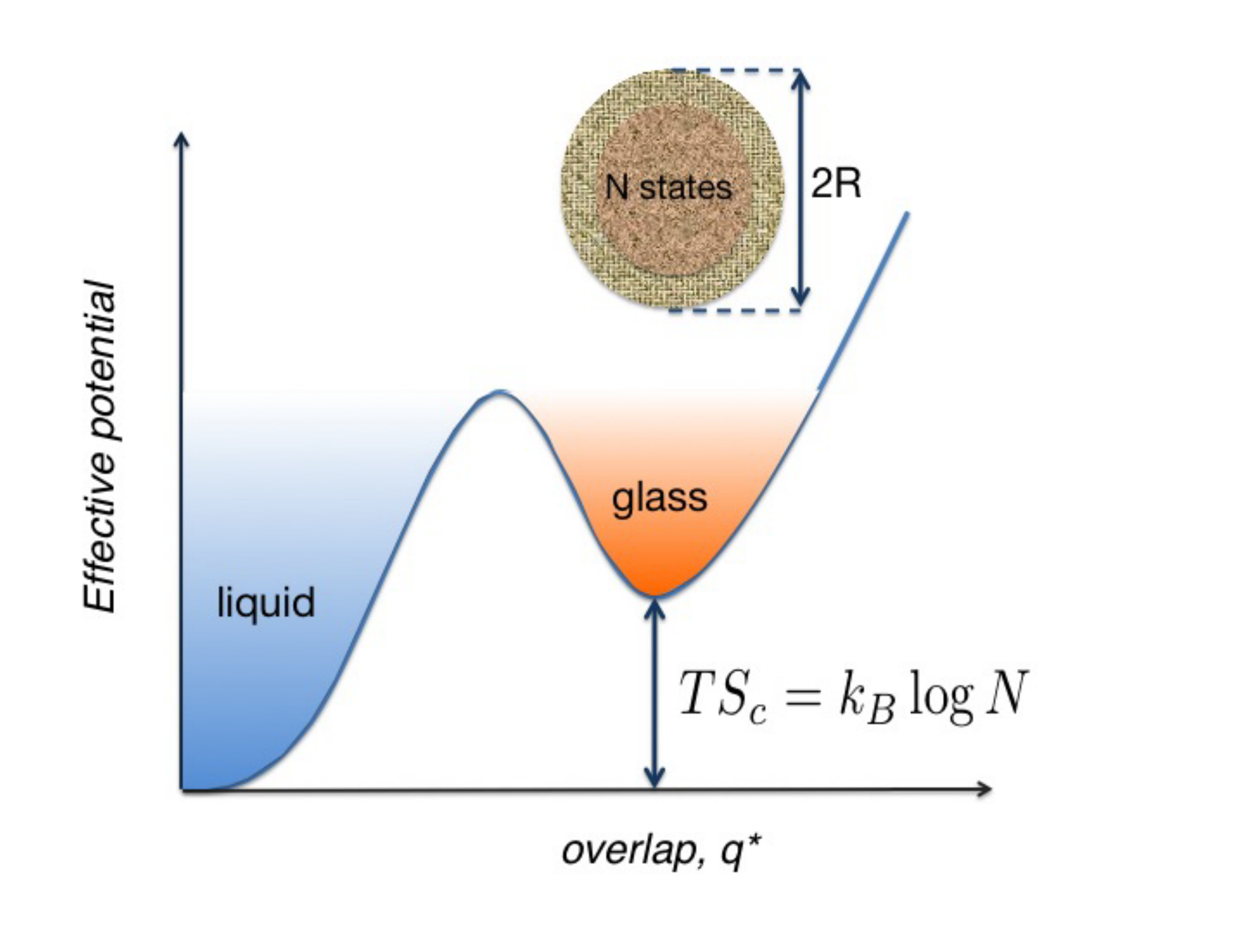}
\caption{Mean-field theory description of the glassy states introduced the
concept of an effective potential. Effective potential is designed to
describe the transition between the non-ergodic glassy states and ergodic
liquid.}
\label{Fig4}
\end{figure}

With $T_{\mathrm{eff}}=T/m$, expressions for $\overline{F}$ and $S_{c}$
follow from the replicated free energy $F\left( T_{\mathrm{eff}}\right) $
via: 
\begin{equation}
\begin{split}
\overline{F} &=-T_{\mathrm{eff}}^{2}\frac{\partial \left( F\left( T_{\mathrm{%
eff}}\right) /T_{\mathrm{eff}}\right) }{\partial T_{\mathrm{eff}}}, \\
S_{c} &=-\frac{\partial F\left( T_{\mathrm{eff}}\right) }{\partial T_{%
\mathrm{eff}}}.  \label{thermo}
\end{split}%
\end{equation}%
These results are in analogy to the usual thermodynamic relations between
free energy ($F\rightarrow F\left( T_{\mathrm{eff}}\right) $), internal
energy ($U\rightarrow \overline{F}$), entropy ($S\rightarrow S_{c}$) and
temperature ($T\rightarrow T_{\mathrm{eff}}$), see also Ref. \cite{N98}. If
the liquid gets frozen in one of the many metastable states, the system
cannot anymore realize its configurational entropy, i.e. the mean free
energy of frozen states is $\overline{F}$, higher by $TS_{c}$ if compared to
the equilibrium free energy of the liquid, Fig. \ref{Fig4}. 

The physically intuitive analogy between effective temperature, mean energy, 
$\overline{F}$ and configurational entropy to \ thermodynamic relations, Eq.
(\ref{thermo}), suggest that one should analyze the corresponding \emph{%
configurational heat capacity} 
\begin{equation}
C_{c}=T_{\mathrm{eff}}\frac{\partial \overline{S_{c}}}{\partial T_{\mathrm{%
eff}}}=-m\frac{\partial \overline{S_{c}}}{\partial m}.
\end{equation}%
Using the distribution function Eq. (\ref{distrms}) we find, as expected,
that $C_{c}$ is\ a measure of the fluctuations of the energy and
configurational entropy of glassy states. \ We obtain for the
configurational heat capacity 
\begin{equation}
C_{c}=\frac{\overline{\left( \delta F\right) ^{2}}}{T_{\mathrm{eff}}^{2}}\ =%
\overline{\left( \delta S_{c}\right) ^{2}}.  \label{ffluctC}
\end{equation}%
where $\overline{\left( \delta F\right) ^{2}}=\overline{F^{2}}-\overline{F}%
^{2}$ and $\overline{\left( \delta S_{c}\right) ^{2}}=\overline{S_{c}^{2}}-%
\overline{S_{c}}^{2}$. Here the mean values $\overline{F}$ and $\overline{%
S_{c}}$ are determined by Eq. (\ref{thermo}). The fluctuations of the
configurational entropy and frozen state energy are then determined by%
\begin{equation}
\overline{S_{c}^{2}}=\int D\psi P\left[ \psi \right] \log ^{2}P\left[ \psi %
\right]
\end{equation}%
and%
\begin{equation}
\overline{F^{2}}=\int D\psi P\left[ \psi \right] F\left[ \psi \right] ^{2},
\end{equation}
respectively. Both quantities can be expressed within the replica formalism
in terms of a second derivative of $F\left( m\right) $ with respect to $m$.
For example it follows: 
\begin{equation}
\frac{\partial ^{2}}{\partial m^{2}}mF\left( m\right) =-\frac{1}{T}\left( 
\overline{F^{2}}-\overline{F}^{2}\right) .  \label{repsecd}
\end{equation}%
It is then easy to show that Eq. (\ref{ffluctC}) holds. With the
introduction of the configurational heat capacity into the formalism we have
a measure for the deviations of the number of metastable states from their
mean value. The analogy of these results to the usual fluctuation theory of
thermodynamic variables \cite{Landau} further suggests that $C_{c}$ also
determines fluctuations of the effective temperature with mean square
deviation: 
\begin{equation}
\overline{\left( \delta T_{\mathrm{eff}}\right) ^{2}}=T_{\mathrm{eff}%
}^{2}/C_{c}.
\end{equation}%
Since $C_{c}$ is extensive, fluctuations of intensive variables, like $T_{%
\mathrm{eff}}$, or densities, like $s_{c}=S_{c}/V$, vanish for infinite
systems. However, they become relevant if one considers finite subsystems or
small droplets. In the context of glasses this aspect was first discussed in
Ref. \cite{Donth}.

If the replica theory is marginally stable, i.e. the lowest eigenvalue of
the fluctuation spectrum beyond mean field solution vanishes, it was shown
in Ref. \cite{WSW03} that $T_{\mathrm{eff}}$ agrees with the result obtained
from the generalized fluctuation-dissipation theorem in the dynamic
description of mean field glasses \cite{CK93}. Typically, the assumption of
marginality is appropriate for early times right after a rapid quench from
high temperatures. \ In this case $T_{\mathrm{eff}}>T$ for $T<T_{A}$, i.e.
the distribution function of the metastable states is not in equilibrium on
the time scales where mode coupling theory or the requirement for marginal
stability applies. Above the Kauzmann temperature it is however possible to
consider the situation where $T_{\mathrm{eff}}=T$, i.e. where the
distribution of metastable states has equilibrated with the external heat
bath of the system. Since we are interested in the restoration of ergodicity
for $T_{K}<T<T_{A}$ we use $T_{\mathrm{eff}}=T$. Below the Kauzmann
temperature the assumption $T_{\mathrm{eff}}=T$ cannot be made any longer as
it implies a negative configurational entropy, inconsistent with the
definition Eq. (\ref{Sc}). It implies that the glass transition is now
inevitable even if one could wait for arbitrarily long times. While this is
likely an idealization of mean field theory, it demonstrates that the ageing
behavior below and above the Kauzman temperature are very distinct.

\subsection{Dynamical interpretation of the replica approach}

Ultimately, glass formation is a dynamic phenomenon. It is therefore useful
and illustrative to offer a dynamical interpretation of the replica
formalism of Ref. \cite{Mon95}. Below we will provide a qualitative set of
arguments. A detailed and quantitative derivation of the formalism, based on
the dynamic theory of glassy systems by Kurchan and Cugliandolo \cite{CK93},
is given in Ref. \cite{WSW04}.

Consider a system characterized by a field variable $\varphi \left( \mathbf{x%
},t\right) $. Let us assume that certain field configurations evolve
extremely slowly, and equilibrate at a much later times than others. In
terms of particle coordinates, a natural choice for slow and fast variables
would be the fiducial position $\mathbf{R}_{i}\left( t\right) $ and the
harmonic vibrations $\mathbf{u}_{i}\left( t\right) $, according to $\mathbf{x%
}_{i}\left( t\right) =$ $\mathbf{R}_{i}\left( t\right) +\mathbf{u}_{i}\left(
t\right) $. In our field theoretic description we assume that slow field
configurations are characterized by $\psi \left( \mathbf{x},t\right) $ while
we continue to refer to fast configurations as\ $\varphi \left( \mathbf{x}%
,t\right) $. Let the dynamics of the fast degrees of freedom be governed by
the Langevin equation 
\begin{equation}
\Gamma \frac{\partial \varphi \left( \mathbf{x},t\right) }{\partial t}=-%
\frac{\delta \mathcal{S}\left( \varphi ,\psi \right) }{\delta \varphi \left(
x,t\right) }+\eta \left( x,t\right) ,
\end{equation}%
where 
\begin{equation}
\mathcal{S}\left( \varphi ,\psi \right) =\mathcal{S}\left[ \varphi \right] \
+\frac{g}{2}\int dx\left( \varphi \left( x\right) -\psi \left( x\right)
\right) ^{2}
\end{equation}%
contains a weak coupling ($g\ll 1$) between fast and slow degrees of
freedom, just like in the replica approach discussed above. Fast variables $%
\ \varphi \left( \mathbf{x},t\right) $ are subject to white noise with
temperature $T$ 
\begin{equation}
\left\langle \eta \left( \mathbf{x},t\right) \eta \left( \mathbf{x}^{\prime
},t^{\prime }\right) \right\rangle =2\Gamma T\delta \left( \mathbf{x}-%
\mathbf{x}^{\prime }\right) \delta \left( t-t^{\prime }\right) .
\end{equation}%
The equilibrium distribution function for fixed $\psi $ is then 
\begin{equation}
p\left[ \varphi |\psi \right] =\frac{1}{Z\left[ \psi \right] }\exp \left(
-\beta \mathcal{S}\left( \varphi ,\psi \right) \right) 
\end{equation}%
where $Z\left[ \psi \right] =\int D\varphi \exp \left( -\beta \mathcal{S}%
\left[ \varphi ,\psi \right] \right) $. \ Next we assume that the dynamics
of the slow degrees of freedom is then governed by 
\begin{equation}
\Gamma _{s}\frac{\partial \psi \left( \mathbf{x},t\right) }{\partial t}=-%
\frac{\delta F\left[ \psi \right] }{\delta \psi \left( x,t\right) }+\eta
_{s}\left( x,t\right) .
\end{equation}%
with $F\left[ \psi \right] =-T\log Z\left[ \psi \right] $. Thus, during the
entire dynamics of the slow degrees of freedom, the fast modes are assumed
to be in equilibrium already. It is important to assume different noise for
the slow variables $\psi \left( x,t\right) $ and for the fast variables $%
\varphi \left( x,t\right) $. \ The variables $\psi \left( \mathbf{x}%
,t\right) $ will, in general, not equilibrate with temperature $T$, but may
be characterized by an effective temperature $T_{\mathrm{eff}}\geq T$, i.e.
we write: 
\begin{equation}
\left\langle \eta _{s}\left( x,t\right) \eta _{s}\left( x^{\prime
},t^{\prime }\right) \right\rangle =2\Gamma _{s}T_{\mathrm{eff}}\delta
\left( x-x^{\prime }\right) \delta \left( t-t^{\prime }\right) 
\end{equation}%
Then, the stationary probability distribution for $\psi $ is 
\begin{equation}
p\left[ \psi \right] =\frac{1}{Z}\exp \left( -\beta _{\mathrm{eff}}F\left[
\psi \right] \right) =\frac{Z\left[ \psi \right] ^{m}}{Z}
\end{equation}%
with $m=T/T_{\mathrm{eff}}$. The joint distribution of $\varphi $ and $\psi $
can now be written as $p\left[ \varphi ,\psi \right] =p\left[ \psi \right] p%
\left[ \varphi |\psi \right] $. It is straightforward to show that this
distribution gives results identical to the replica approach. Above the
Kauzmann temperature, $T_{K}$, equilibration of the slow variables is still
possible and $T$ approaches $T_{\mathrm{eff}}$, i.e. $m\rightarrow 1$

\section{Glass formation in uniformly frustrated systems}

To gain detailed insight into the possibility of glass formation, we now go
back to our model described by and analyze the problem (\ref{Hint}) with $%
W\left( \varphi \right) =\frac{u_{4}}{4}\varphi^{4}$ analytically within the
self consistent screening approximation (SCSA) \cite{SCSA}, an approximate
treatment that is controlled by an expansion in $1/N$, where $N$ is the
number of components of $\varphi $, see Ref. \cite{SWW00}.

The free energy $F(m)$ of the replicated Hamiltonian is given in terms
of the regular correlation function $G(q)$ and the correlation
function $%
F(q)\equiv \langle \varphi ^{a}(q)\varphi ^{b}(-q)\rangle $ for $a\neq
b$, i.e. between the fields in different replicas. The latter
corresponds to the Edward-Anderson parameter signaling a glassy
state. For a system in the universality class of the random first
order transition we expect that below a temperature $T_{A\text{ }}$
the system establishes an exponentially large number of metastable
states and long time correlations, characterized by the correlation
function $F\left( \mathbf{x,x}^{\prime }\right)
=T^{-1}\lim_{t\rightarrow \infty }\left\langle \varphi \left(
    \mathbf{x}%
    ,t\right) \varphi \left( \mathbf{x}^{\prime },0\right)
\right\rangle $.  These long time correlations occur even though no
state with actual long range spatial order exists.

Within the SCSA, the relevant part of the free energy $F(m)$ is: 
\begin{equation}
F(m)=-\frac{T}{m}(\mathrm{Tr}\ln \mathcal{G}^{-1}+\mathrm{Tr}\ln \mathcal{D}%
^{-1}),  \label{fe}
\end{equation}%
which determines the configurational entropy $S_{c}=\left. \frac{1}{T}\frac{d%
\mathcal{F}(m)}{dm}\right\vert _{m=1}$. Here, 
\begin{equation}
\mathcal{G}\equiv (G-F)\mathbf{I}+F\mathbf{E}
\end{equation}%
is the correlation function matrix with $\mathbf{I}_{ab}=\delta _{ab}$ and $%
\mathbf{E}_{ab}=1$. The symbol \textrm{Tr} in Eq.~(\ref{fe}) includes the
trace of the replica space and the momentum integration. The matrix $%
\mathcal{D}$ is related to $\mathcal{G}$ via 
\begin{equation}
\mathcal{D}^{-1}=(uT)^{-1}\mathbf{I}+\Pi ,
\end{equation}%
where 
\begin{equation}
\Pi =(G\otimes G-F\otimes F)\mathbf{I}+(F\otimes F)\mathbf{E}
\end{equation}%
is the generalized polarization matrix. The symbol $\otimes $ denotes a
convolution in Fourier space. The replicated Schwinger-Dyson equation can be
written as%
\begin{equation}
\mathcal{G}^{-1}=G_{H}^{-1}\mathbf{I}+\Sigma ,  \label{Dyson}
\end{equation}
where $\Sigma $ is the self-energy matrix and $G_{H}\left( k\right) $ the
Hartree correlation function of Eq. (\ref{Braz}).

Within the SCSA the self-energy has diagonal elements $\Sigma _{G}=2G\otimes
D_{G}$ and off diagonal elements $\Sigma _{F}=2F\otimes D_{F}$ in replica
space, where $D_{G}$ and $D_{F}$ being, respectively the diagonal and
off-diagonal elements of $\mathcal{D}$. These equations form a closed set of
self-consistent equations which enable us to solve for $G$ and $F$, and then
determine the configurational entropy. To evaluate $F(m)$ we use the fact
that a matrix of the form $A_{ab}=\left( a-b\right) \delta _{ab}+b$ has $m-1$
degenerate eigenvalues $a_{i}=a-b$ where $i=1,m-1$ and and one eigenvalue $%
a_{m}=a+\left( m-1\right) b$. This yields 
\begin{equation}
\mathrm{Tr}\log A=\sum_{i=1}^{m}\int \frac{d^{d}k}{\left( 2\pi \right) ^{d}}%
\log a_{i}
\end{equation}
This result allows us to evaluate $F\left( m\right) $ of Eq. (\ref{fe}).
Performing the derivative of the resulting expression with respect to $m$
then yields the configurational entropy 
\begin{equation}
S_{\mathrm{c}}=\int \frac{d^{d}q}{\left( 2\pi \right) ^{d}}\left\{ s\left[ 
\frac{F}{G}\right] -s\left[ \frac{F\otimes F}{(u_{4}T)^{-1}+G\otimes G}%
\right] \right\} ,  \label{Sconf}
\end{equation}%
were, $s[x]\equiv -x-\ln (1-x)$.

The analysis of these coupled equations reveals that the self energies
$%
\Sigma _{G}$ and $\Sigma _{F}$ are only weakly momentum
dependent. Then the impact of $\Sigma _{G}$ is solely to renormalize
the correlation length $\xi $. However, the emergence of $\Sigma _{F}$
leads to a qualitative change.  Dimensional analysis reveals that
$\Sigma _{F}$ is an inverse length squared, which motivates one to
introduce the Lindemann length $\lambda $ of the glass via $\lambda
^{-2}=-A_{0}\Sigma _{F}$. An interpretation of this length scale in
terms of slow defect motion in a stripe glasses was given in
Ref. \cite{WSW03}. By inspection of the Dyson equation and using the
fact that $G\left( k\right) $ is strongly peaked at the modulation
wave number $%
k_{0}$ one finds for $k\sim k_{0}$ that
\begin{equation}
F\left( k_{0}\right) \lesssim G\left( k_{0}\right) .
\end{equation}%
$G\left( k\right) $ vanishes rapidly away from the peak (as does $\Sigma
_{F}(k_{0})G\left( k_{0}\right) )$) and it follows from the same equation, %
\ref{Dyson}, that for large $\left\vert k-k_{0}\right\vert $ holds:$\ $ 
\begin{equation}
F\left( k\right) \simeq -\Sigma _{F}\left( k_{0}\right) G^{2}\left( k\right).
\end{equation}%
If a solution for $F$ exists, it is going to be a peaked at $k_{0}$,
but smaller and narrower than $G$. Consequently, if a stripe glass
occurs, the long time limit of the correlation function is not just a
slightly rescaled version of the instantaneous correlation function,
but it is multiplied by a $k$ dependent function that leads to a
qualitatively different behavior for different momenta. Once a glassy
state is formed, configurations which contribute to the peaks of
$G\left( k\right) $ and $F\left( k\right) $, i.e.  almost perfect
stripe configurations, are almost unchanged even after long
times. Close to $k_{0}$, $F\left( k\right) \,$\ is solely reduced by
some momentum independent Debye-Waller factor $\exp \left( -D\right)
=F\left( k_{0}\right) /G\left( k_{0}\right) $. On the other hand,
configurations which form the tails of $G\left( k\right) $,
i.e. defects and imperfections of the stripe pattern, disappear after
a long time since now $F\left( k\right) \ll G\left( k\right) $. The
ratio of both functions is now strongly momentum dependent. $F\left(
  k\right)$ becomes sharper than $G\left( k\right) $ because local
defects got healed in time. The length scale that determines the
transition between these two regimes is the length $\lambda $%
. \ This length can therefore be associated with the allowed
generalized vibrational motions in a potential minima of the complex
energy landscape of the system. In analogy with structural glasses we
therefore call $\lambda $ the Lindemann length of the stripe glass.
\begin{figure}[h]
\includegraphics[scale=0.28,angle=0]{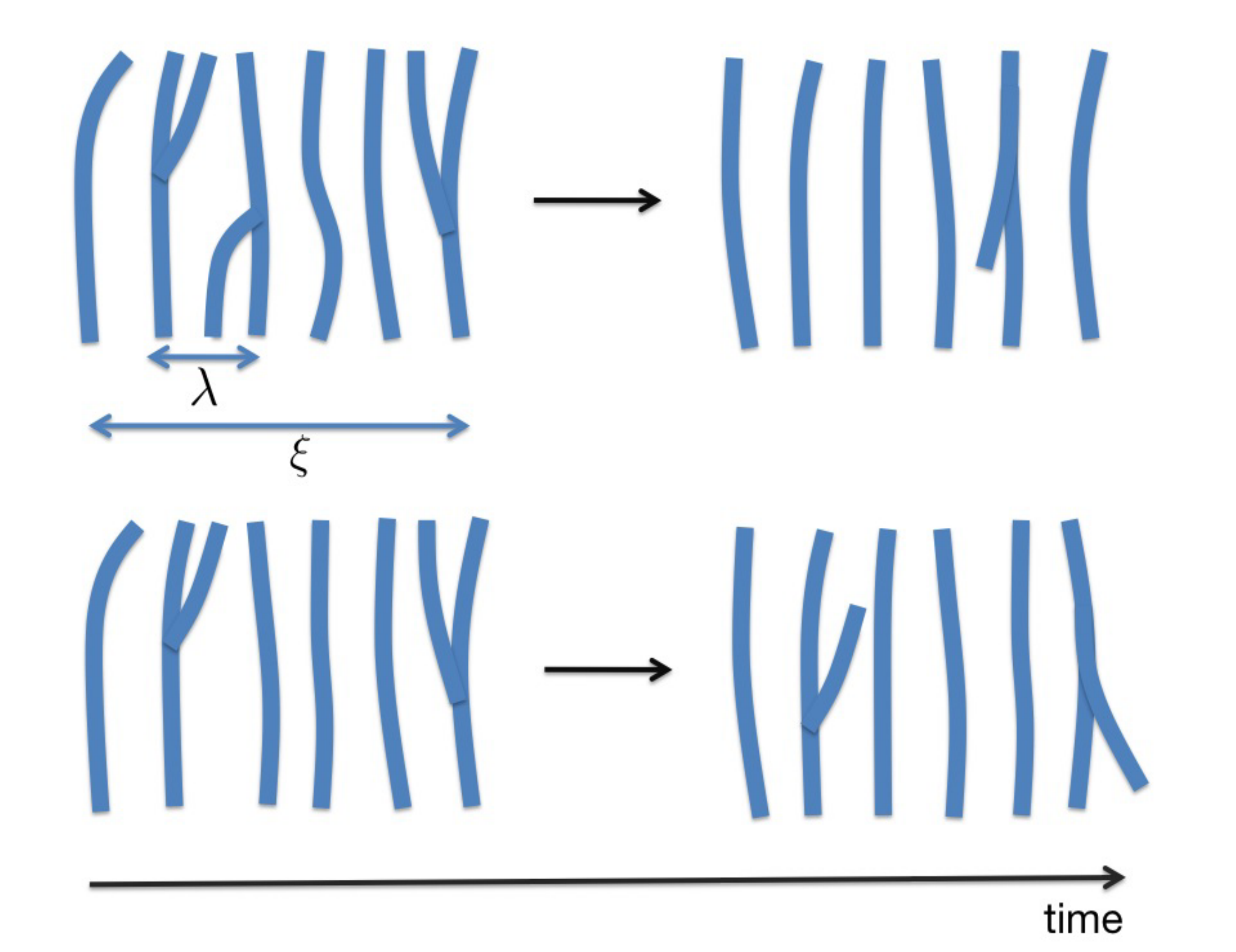}
\caption{Schematic presentation of the defects wandering in structural
glasses. Here $\protect\lambda$ is an average distance over which the
defects are manage to wander (Lindemann length) and $\protect\xi$ is the
correlation length (see text). Upper panel illustrates the situation when
defects are relatively close to each other so that they disappear with time
and system effectively heals itself. Bottom panel depict the opposite
situation when wandering of the defects does not produce healing since they
are on distances $l>\protect\lambda$ from each other (see Ref. \protect\cite%
{SWW00} for details.)}
\label{Fig5}
\end{figure}

\begin{figure}[h]
\includegraphics[scale=0.28,angle=0]{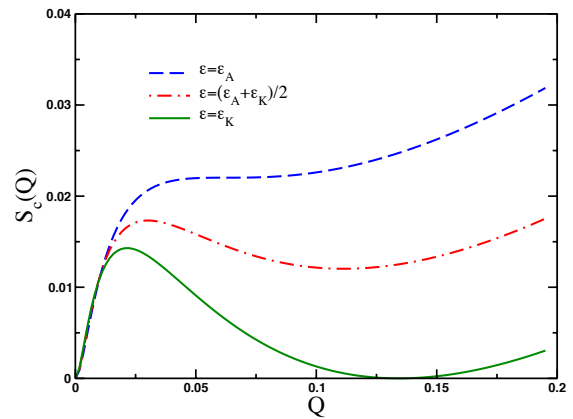}
\caption{Plot of the configurational entropy $S_c(Q)$ as a function of the
dimensionless parameter $Q=(1/\protect\lambda q_0)^2$, Eq. (4.10) for
different values of another parameter $\protect\varepsilon=\protect\xi q_0$.
It is instructive to compare this plot with the schematic plot of Fig. 4.}
\label{Fig6}
\end{figure}

For our subsequent analysis we introduce the dimensionless quantity $Q$ via 
\begin{equation}  \label{SigmaF}
\Sigma _{F}=-Q\frac{q_{0}^{2}}{A_{0}},
\end{equation}%
i.e. $\lambda ^{-2}=Qq_{0}^{2}$. We are now in a position to perform
the momentum integration and evaluate the configurational entropy as
function of $Q$. It follows that
\begin{equation}
S_{c}\left( Q\right) =Vk_{0}^{d}K_{d}\left[ \frac{1}{\xi q_0}\frac{z^{2}}{%
(1-z)}+\frac{2}{\pi }\left( z^{2}+\log \left( 1-z^{2}\right) \right) \right]
\label{SofQ}
\end{equation}%
where%
\begin{equation}
z\left( Q\right) =1-\frac{1}{\sqrt{1+Q\xi ^{2}q_{0}^{2}}}.
\end{equation}%
In Fig. \ref{Fig6} we show our results for $S_{c}\left( Q\right) $ for
different values of $\varepsilon \equiv \xi q_{0}$. A self consistent
solution of the above set of coupled equations corresponds to finding
stationary points $Q^{\ast }$ with 
\begin{equation}
\left. \partial S_{c}/\partial Q\right\vert _{Q=Q^{\ast }}=0.
\label{sstationary}
\end{equation}%
$Q^{\ast }=0$ is always a solution with $S_{c}\left( 0\right) =0$. In
addition there is a locally stable solution for $\varepsilon
>\varepsilon _{A}=11.770574$ with configurational entropy given as
$S_{c}\left( Q_{A}^{\ast }\right) =Vk_{0}^{d}K_{d}\times 0.0220247$
and $Q_{A}^{\ast }=Q^{\ast }\left( \varepsilon _{A}\right)
=0.0625528$. The configurational entropy at the metastable maximum
vanishes at $\varepsilon =\varepsilon _{K}=13.169625$ with
$Q_{K}^{\ast }=0.134271$. \ This is the behavior that is generally
expected in a system a with random first order transition. Upon
lowering the temperature the ergodic, liquid like state becomes more
and more correlated, i.e. the correlation length growths. Once $\xi $
reaches a threshold value $\xi _{A}=\frac{\varepsilon _{A}}{2\pi
}l_{0}\simeq 1.\,\allowbreak 87l_{0}$ about twice the typical length
scale for field modulations $l_{0}=2\pi /q_{0}$, glassy dynamics sets
in as consequence of an emergence of exponentially many states. The
Kauzman temperature is reached once the liquid state correlation
length grows even further and becomes equal to $\xi
_{K}=\frac{\varepsilon _{K}}{2\pi }l_{0}\simeq 2.1l_{0}
$. 

It is also interesting to analyze the penalty for a spatially varying
overlap $Q\to Q({\mathbf{x}})$. This will be essential for the Landau
expansion and the droplet arguments that follow below. Performing a
gradient expansion of the replica free energy yields the correction
\begin{equation}
\delta F(m)=-\frac{a_0^2}{2}(m-1)T\int d^dx\left({\vec\nabla}Q\right)^2,
\end{equation}
where 
\begin{equation}  \label{a0}
a_0^2=\frac{k_0^4}{A_0^2}\lim\limits_{q\to 0}\frac{1}{q^2}\int\frac{d^dk}{%
(2\pi)^d}\left[G(k)G(k+q)-G^2(k)\right]
\end{equation}
The prefactor $k_0^4/A_0^2$ appears from our definition for $Q$ (\ref{SigmaF}%
). For $d=3$ and $\tau=1$ the integration can be performed exactly with the
following result \cite{SWW00}: 
\begin{equation}  \label{a02}
a_0^2=\frac{k_0}{48\pi}(\xi k_0)^5.
\end{equation}
As was anticipated in Ref. \cite{SWW00}, the penalty for a spatial
deformation becomes small as $k_0\to 0$, i.e. in the limit where the
modulation length is very large. Note that at $T=T_A$ or $T=T_K$, the
dimensionless product $\xi k_0$ is independent of $k_0$, i.e. varying
the characteristic scale yields $a_0(T_A)\propto k_0^{1/2}$.

Thus, we analyzed an analytically treatable example for a random first order
transition which enabled us to identify the underlying physical mechanism
for glassiness in a uniformly frustrated system. Once equilibrium
correlations are sufficiently strong and the stripe liquid - stripe solid
transition is kinematically inaccessible, the system is governed by slow
glassy dynamics. At this point a Lindemann length, $\lambda $, emerges,
which is a length scale over which imperfections of the stripe pattern
manage to wander, Fig. \ref{Fig5}. Defects of the perfectly ordered state
are still abundant as the healing length is finite, i.e. the system enters a
random solid state. Local order is however established. This may not be
identical to the most stable crystalline, i.e. long ranged ordered,
configuration, but rather correspond to configurations that are occurring
with high weight close to the spinodal of the system. In the context of
supercooled liquids, the natural analog of such locally correlated
configuration are icosahedral or body centered cubic short range order.

Computer simulations on the model discussed above with $u_{3}=0$,
suggest that the nucleation barriers of the crystalline state are
rather low \cite%
{Geissler04}. However, the inclusion of asymmetric terms in the
interaction potential ($u_{3}\neq 0$) will likely strengthen first
order transitions to a crystalline state and enhance the nucleation
barriers. This behavior was analyzed in
Refs. \cite{WSW04,Zhang06,Wu09}.

\section{Replica Landau Theory}

The analysis of the previous section demonstrated that \ uniformly
frustrated systems undergo self generated glass transitions with an entropy
crisis, of the random first order transition universality class. These
calculations are on the mean field level and ignore issues related to
dynamical heterogeneity, i.e. variations in the typical time scales in
spatially nearby regions. The relevance of such effect is already evident
from the fact that fluctuations in the heat capacity become important in
finite regions, as shown above. To capture these phenomena we need to go
beyond the mean field level and include droplet formations. Droplets are not
regions with different values of the field $\varphi \left( \mathbf{r}\right) 
$ that underlies our analysis. Instead, by droplets we mean a spatial
variation of the Lindemann length, i.e. of the parameter $Q$ used above to
characterize the overlap between configurations at distant times. Thus, it
is useful to consider instead of $Q$ a dynamic field $Q_{ab}\left( \mathbf{r}%
\right) $ that generates the replica partition function, i.e. 
\begin{equation}
Z\left( m\right) =\int DQ\exp \left( -\beta H_{\mathrm{eff}}\left[ Q\right]
\right) .
\end{equation}%
The mean field solution of this problem is then determined by $\delta H_{%
\mathrm{eff}}\left[ Q\right] /\delta Q_{ab}=0$ with the Ansatz 
\begin{equation}
Q_{ab}=Q^{\ast }\left( 1-\delta _{ab}\right)   \label{mf}
\end{equation}%
with $Q^{\ast }$ playing the exact same role as the corresponding quantity
determined by the condition Eq. (\ref{sstationary}) in the previous section.

Formally $H_{\mathrm{eff}}\left[ Q\right] $ can be obtained from the field
theory discussed above by multiplying $Z\left( m\right) $ of Eq. (\ref{Zmm})
by unity%
\begin{equation}
\begin{split}
1=& \int D\mu \prod\limits_{\mathbf{x,x}^{\prime },ab}\delta \left( \mu
_{ab}\left( \mathbf{x,x}^{\prime }\right) -\varphi ^{a}(\mathbf{x})\varphi
^{b}(\mathbf{x}^{\prime })\right) = \\
=& \int\limits_{-i\infty }^{i\infty }DQ\int D\mu \exp \left\{ -\int_{\mathbf{%
x,x}^{\prime }}\sum_{ab}Q_{ab}\left( \mathbf{x,x}^{\prime }\right) \times
\right. \\
& \left. \times \left[ \mu _{ab}\left( \mathbf{x,x}^{\prime }\right)
-\varphi ^{a}(\mathbf{x})\varphi ^{b}(\mathbf{x}^{\prime })\right] \right\}
\end{split}%
\end{equation}%
and switch the order of integration. Then we find 
\begin{equation}
\begin{split}
& e^{-H_{\mathrm{eff}}\left[ Q\right] }=\int D\mu D^{m}\varphi
e^{-\sum_{a=1}^{m}\beta \mathcal{H}\left[ \varphi ^{a}\right] }\times \\
& \times e^{-\sum_{a,b=1}^{m}\int_{\mathbf{x,x}^{\prime }}Q_{ab}(\mathbf{x,x}%
^{\prime })\left( \mu _{ab}(\mathbf{x,x}^{\prime })-\varphi ^{a}(\mathbf{x}%
)\varphi ^{b}(\mathbf{x}^{\prime })\right) }.
\end{split}%
\end{equation}%
We consider the Fourier transform of $Q_{ab}\left( \mathbf{x,x}^{\prime
}\right) $ with respect to the relative coordinates $\mathbf{x-x}^{\prime }$
and approximate it by its value at $k=k_{0}$. This is consistent with our
finding that the self energies of the mean field theory depend only weakly
on momentum. We do however include the dependence on the center of gravity
coordinate to analyze spatially varying overlaps between distant
configurations, yielding $Q_{ab}\left( \mathbf{x}\right) $.

In order to keep our calculation transparent we will not analyze the model
discussed for uniformly frustrated systems, but instead start from a simpler
Landau theory that is in the same universality class \cite{Gross85,Dzero05}: 
\begin{equation}
H_{\mathrm{eff}}\left[ Q\right] =E_{0}\sum_{a,b}\int \frac{d^{3}r}{a_{0}^{3}}%
\left( h\left[ Q_{ab}\right] -\frac{u}{3}\sum_{c}Q_{ab}Q_{bc}Q_{ca}\ \right)
\label{ham1}
\end{equation}%
with 
\begin{equation}
h\left[ Q_{ab}\right] =\frac{a_{0}^{2}}{2}\left( \nabla Q_{ab}\right) ^{2}+%
\frac{t}{2}Q_{ab}^{2}-\frac{u+w}{3}Q_{ab}^{3}+\frac{y}{4}Q_{ab}^{4}
\label{ham2}
\end{equation}%
and replica index $a$,$b=1,\ldots ,m$. $\ a_{0}$ is a length scale of
the order of the first peak in the radial distribution function of the
liquid and $E_{0}$ is a typical energy of the problem that determines
the absolute value of $T\overline{s_{c}}$. In addition the problem is
determined by the dimensionless variables $t$, $u$, $w$ and $y$, which
are in principle all temperature dependent. We assume that the primary
$T$-dependence is that of the quadratic term, where
$t=\frac{T-T_{0}}{E_{0}}$. In what follows we further simplify the
notation and measure all energies in units of $E_{0}$ and all length
scales in units of $a_{0}$. Formally, Eqs. (\ref{ham1}) \ and
(\ref{ham2}) can be motivated as the Taylor expansion of \ the free
energy $F%
\left[ m\right] $ with respect to $\Sigma _{ab}$. \ In practice, the
explicit Taylor expansions of more microscopic models do not yield
quantitatively reliable results. For example, the Taylor expansion of
Eq. (%
\ref{SofQ}) poorly reproduces the behavior in the glassy
regime. Still, the simple Landau expansion of Eqs. (\ref{ham1}) \ and
(\ref{ham2}) allows us to gain key insights that enable us to
investigate the more general case. In particular, we will demonstrate
that the replica instanton theory can easily be generalized to the
more complex stripe glass model.

The mean field analysis of the model Hamiltonian Eqs. (\ref{ham1}) and (\ref%
{ham2}) is straightforward. Inserting the Ansatz Eq. (\ref{mf}) into $H_{%
\mathrm{eff}}\left[ Q\right] $ and minimizing w.r.t $Q$ yields $Q^{\ast }=0$
or 
\begin{equation}
Q^{\ast }\left( t\right) =\ \frac{w+\sqrt{w^{2}-4ty}}{2y}.  \label{qstar}
\end{equation}%
Nontrivial solutions exits for $t<t_{A}=\frac{w^{2}}{4y}$ with $Q_{A}^{\ast
}=Q^{\ast }\left( t_{A}\right) =\frac{w}{2y}$, which determines the mode
coupling temperature $T_{A}=t_{A}E_{0}+T_{0}$. Inserting $Q^{\ast }$ of Eq. (%
\ref{qstar}) into $H_{\mathrm{eff}}\left[ Q\right] $ yields at the saddle
point $F\left( m\right) =H_{\mathrm{eff}}\left[ Q\right] $ for the replica
free energy, which yields: 
\begin{equation}
\frac{F\left( m\right) }{V\left( m-1\right) }=\frac{t}{2}Q^{\ast 2}-\frac{%
w+u\left( m-1\right) }{3}Q^{\ast 3}+\frac{y}{4}Q^{\ast 4}.  \label{F(m)Mon}
\end{equation}%
\ The configurational entropy, as determined by $S_{c}=\left. \frac{1}{T}%
\frac{\partial }{\partial m}F\left( m\right) \right\vert _{m\rightarrow 1}$
is given by 
\begin{equation}
S_{c}\left( Q\right) =\frac{V}{T}\left( \frac{t}{2}Q^{2}-\frac{w}{3}Q^{3}+%
\frac{y}{4}Q^{4}\right) .  \label{Scmon}
\end{equation}%
The stationary points of $S_{c}\left( Q\right) $ are given by $Q^{\ast }$ of
Eq.\ref{qstar}. Inserting $Q^{\ast }$ yields that $S_{c}$ vanishes at $t_{K}=%
\frac{2w^{2}}{9y}$ with $Q_{K}^{\ast }=\frac{2w}{3y}$. Close to $t_{K}$ it
follows that 
\begin{equation}
S_{c}\simeq V\frac{t_{K}}{y}\left( \frac{t-t_{K}}{T_{K}}\right) \propto V%
\frac{T-T_{K}}{T_{K}}
\end{equation}%
as expected. At $t_{A}$ one finds $T_{A}S_{c}\left( T_{A}\right) =V\frac{%
w^{4}}{192y^{3}}$. $\ \ $For the configurational heat capacity follows%
\begin{equation}
C_{c}\ =\frac{2V}{T}\ \left( \frac{t}{2}Q^{\ast 2}-\frac{w+u}{3}Q^{\ast 3}+%
\frac{y}{4}Q^{\ast 4}\right) 
\end{equation}%
It holds at $C_{c}\left( T_{K}\right) =\frac{Vu}{T_{K}}\frac{2}{3}\left( 
\frac{2w}{3y}\right) ^{3}$ and we can write%
\begin{equation}
S_{c}\simeq \frac{w}{4u}C_{c}\left( t_{K}\right) \ \frac{t-t_{K}}{t_{K}}.
\end{equation}%
Thus, we see that the main findings of the model calculations are reproduced
by the simple Landau expansion, Eqs. (\ref{ham1}) and (\ref{ham2}). They
will now be used as starting point for our analysis of dynamical
heterogeneity in form of a replica instanton theory.

\bigskip

\section{Replica instantons and entropic droplets}

At the mean field level a glass at $T<T_{A}$ is frozen in one of many
metastable states. $Q^{\ast }$ then characterizes the overlap between
configurations at distant times. The free energy of such a frozen state is
higher by $TS_{c}$ compared to the ergodic liquid state that is
characterized by $Q^{\ast }=0$. Thus, for $T_{K}<T<T_{A}$ the mean field
glass is locally stable. Local stability also follows from the fact that the
lowest eigenvalue of the fluctuation matrix $\delta ^{2}H/\delta
Q_{ab}\left( \mathbf{x}\right) \delta Q_{cd}\left( \mathbf{x}^{\prime
}\right) $ is positive for $T_{K}\leq T<T_{A} $ and vanishes at $T=T_{A}$,
see Ref. \cite{Dzero05}. The $Q$-dependence of $S_{c}\left( Q\right) $ shown
in Fig. \ref{Fig6} suggests that the decay modes for the frozen state are
droplet excitations, similar to the nucleation of an unstable phase close to
a first order transition. This situation was analyzed in \ Ref. \cite%
{Franz05-1,Dzero05}. In agreement with the RFOT theory \cite{KTW89}, the
driving force for nucleation is the configurational entropy, leading to the
notion of entropic droplets. The formal approach to analyze entropic
droplets is performed in terms of the effective potential approach of Refs. 
\cite{Franz95,Franz97,Barrat97}. We used this technique to formulate a
replica instanton and barrier fluctuation theory in Refs. \cite%
{Dzero05,Dzero09}. In what follows we use a slightly simpler approach that
yields essentially the same results, but is physically significantly more
transparent.

As can be motivated by the more general effective potential approach of
Refs. \cite{Franz95,Franz97,Barrat97}, instanton solutions for entropic
droplets can be determined from%
\begin{equation}
\frac{\delta H\left[ Q\right] }{\delta Q_{ab}\left( \mathbf{x}\right) }=0,
\label{inststat}
\end{equation}%
where we allow for spatial variations of the overlap $Q_{ab}\left( \mathbf{r}%
\right) =Q\left( \mathbf{r}\right) \left( \delta _{ab}-1\right) $. This
yields the following nonlinear equation for $Q\left( \mathbf{r}\right) $: 
\begin{equation}
\nabla ^{2}Q\left( \mathbf{r}\right) =\frac{ds_{c}\left( Q\left( \mathbf{r}%
\right) \right) }{dQ\left( \mathbf{r}\right) },  \label{inst1}
\end{equation}%
with configurational entropy density $s_{c}\left( Q\right) =S_{c}\left(
Q\right) /V$. In case of the Landau expansion we use $S_{c}\left( Q\right) $
of Eq. (\ref{Scmon}), while for the stripe glass approach we start from Eq. (%
\ref{SofQ}). We will first analyze the simpler case of the Landau theory.
Then Eq. (\ref{inst1}) admits an exact solution in the \ thin wall limit $%
R\gg \xi $: 
\begin{equation}
Q(x)=q^{\ast }+\sqrt{\frac{2}{y\xi ^{2}}}\left[ \text{th}\left( \frac{x}{\xi 
}-z_{0}\right) -\text{th}\left( \frac{x}{\xi }+z_{0}\right) \right] ,
\label{pcthinwall}
\end{equation}%
where the integration constant $z_{0}$ is a function of $t,w$ and $y$. $R$
is the droplet radius and $\xi $ is the interface width given by 
\begin{equation}
\xi =\frac{4a_{0}}{\sqrt{3y(2Q^{\ast }-Q_{K}^{\ast })^{2}-6t_{K}+4t}}.
\end{equation}%
Inserting the solution Eq. (\ref{pcthinwall}) into the expression into $H%
\left[Q\right] $ we calculate the value of the mean barrier. The latter is
determined by optimizing the energy gain due to creation of a droplet and
energy loss due to the surface formation. As a result for the mean barrier
we find (reintroducing the energy scale $E_{0}$ and length scale $a_{0}$) 
\begin{equation}
\overline{F^{\ddagger }}=E_{0}\frac{32\pi a_{0}}{9y\xi ^{3}}R^{2},
\label{Fdagger}
\end{equation}%
The droplet radius 
\begin{equation}
R=\frac{64a_{0}^{4}}{3y^{2}q^{\ast 3}(q_{K}^{\ast }-q^{\ast }\left( t\right)
)\xi ^{3}}
\end{equation}%
is determined from the balance between the interface tension and the
entropic driving force for nucleation. Furthermore, $q_{K}^{\ast }\equiv
q^{\ast }(t=t_{K})$ is the order parameter at the Kauzmann temperature. When
temperature approaches the $T_{K}$ the radius of the droplet as well as the
mean barrier diverge. One finds $\lim\limits_{t\rightarrow t_{K}}\overline{%
F^{\ddagger }}\propto (t-t_{K})^{-2}$ and $\lim\limits_{t\rightarrow
t_{K}}R\propto (t-t_{K})^{-1}$. Since the droplet interface $\xi $ remains
finite as $t\rightarrow t_{K}$, the thin wall approximation is well
justified close to the Kauzman temperature. On the other hand, $R$ and $\xi $
become comparable for temperatures close to $T_{A}$ and the thin wall
approximation breaks down. Combining $R\propto (t-t_{K})^{-1}$ and $%
s_{c}\propto (t-t_{K})$, we obtain $\nu =1$ for the exponent that relates
the droplet size $R$ and the configurational entropy density: $R\propto
s_{c}^{-\nu }$.

Close to $t_{K}$ follows \ $\sigma \left( t\right) \simeq \sigma _{K}\left(
1-\frac{21}{\ 2}\left( \frac{t-t_{K}}{t_{K}}\right) \right) $ with $\sigma
_{K}=\frac{4a_{0}}{27\sqrt{15}}\frac{w^{3}}{y^{5/2}}.$ The critical droplet
nucleation radius is $R=2\sigma /s_{c}$ yielding the barrier 
\begin{equation}
\overline{F^{\ddagger }}=\frac{16\pi \sigma ^{3}}{3s_{c}^{2}}
\end{equation}%
for the nucleation of entropic droplets. This leads to a mean relaxation
time of 
\begin{equation}
\overline{\tau }=\tau _{0}\exp \left( \frac{\overline{F^{\ddagger }}}{k_{B}T}%
\right) .  \label{meanr}
\end{equation}

It was pointed out in Ref. \cite{KTW89} that wetting effects of the
interface alter the a relationship\ between droplet size and entropic
driving force to $R\propto s_{c}^{-\nu }$ with exponent $\nu $. A
renormalization of the droplet interface due to wetting of intermediate
states on the droplet surface was shown to yield $\nu =2/d$ \cite{KTW89},
leading to $\overline{F^{\ddagger }}\propto Ts_{c}^{-1}$ and correspondingly
to a Vogel-Fulcher law 
\begin{equation}  \label{taufrag}
\overline{\tau }=\tau _{0}\exp \left( \frac{DT_{K}}{T-T_{K}}\right)
\end{equation}
for the mean relaxation time$.$

It is straightforward to analyze the more complex configurational entropy of
the stripe glass problem. In this case the nonlinear instanton equation
cannot be solved analytically. However we can make progress by performing a
variational calculation for the droplet. The mean barrier is 
\begin{equation}
\overline{F^{\ddagger }}=H_{\mathrm{eff}}\left[ Q\left( \mathbf{x}\right) %
\right] -S_{c}\left( Q^{\ast }\right) ,
\end{equation}%
where $Q\left( \mathbf{x}\right) $ is a localized instanton solution which
differs from $Q^{\ast }$ in a finite region. We make the trial ansatz:

\begin{equation}
Q\left( \mathbf{x}\right) =\left\{ 
\begin{array}{cc}
0 & \left\vert \mathbf{x}\right\vert <R \\ 
Q^{\ast }\frac{\left\vert \mathbf{x}\right\vert -R}{l} & R<\left\vert 
\mathbf{x}\right\vert <R+l \\ 
Q^{\ast } & \left\vert \mathbf{x}\right\vert >R+l%
\end{array}%
\right.  \label{ansatz}
\end{equation}%
and insert it into the above expression for $H_{\mathrm{eff}}\left[ Q\left( 
\mathbf{x}\right) \right] $. We find 
\begin{equation}
\overline{F^{\ddagger }}\left( R\right) =4\pi \sigma R^{2}-\frac{4\pi }{3}%
s_{c}\left( Q^{\ast }\right) R^{3}
\end{equation}%
with surface tension 
\begin{equation}
\sigma =\frac{a_{0}^{2}Q^{\ast 2}}{2l}+lT\left[ \int_{0}^{Q^{\ast
}}s_{c}\left( Q\right) \frac{dQ}{Q^{\ast }}-s_{c}\left( Q^{\ast }\right) %
\right]
\end{equation}%
Minimizing with respect to the droplet wall thickness $l$ yields%
\begin{equation}
l^{2}=\frac{1}{2}\frac{a_{0}^{2}Q^{\ast 3}}{\int_{0}^{Q^{\ast }}s_{c}\left(
Q\right) dQ-s_{c}\left( Q^{\ast }\right) Q^{\ast }}
\end{equation}%
and correspondingly for the surface tension 
\begin{equation}
\sigma =\sqrt{2}a_{0}Q^{\ast 1/2}\left( \int_{0}^{Q^{\ast }}s_{c}\left(
Q\right) dQ-s_{c}\left( Q^{\ast }\right) Q^{\ast }\right) ^{1/2}
\end{equation}%
Using the our result Eq. (\ref{SofQ}) for the configurational entropy as a
function of $Q$ for the stripe glass problem in $d=3$ and for $\tau =1$
(i.e. with the long range Coulomb interaction $V({\mathbf{x}})$) and Eq. (%
\ref{a02}) for the gradient coefficient $a_{0}$, yields at $T_{K}$: 
\begin{equation}
\sigma (T_{K})=C_{\sigma }T_{K}k_{0}^{2},
\end{equation}%
where $C_{\sigma }\simeq 3.45\times 10^{-2}$. Thus we see that the surface
tension of entropic droplets vanishes in the limit $k_{0}\rightarrow 0$. For
the wall thickness we obtain $l(T_{K})\simeq k_{0}^{-1}$.

Finally, we use our results for the surface tension to analyze the variation
of the fragility $D$ in Eq. (\ref{taufrag}) as function of $k_0$. One finds
\begin{equation}
D=\frac{3\sigma^2}{T_K^3\left(\frac{dS_c}{dT}\right)_{T=T_K}}
\end{equation}
which was further investigated in a numerical analysis of the stripe
glass problem in Ref. \cite{Grousson2001}. This analysis led to
$D(k_0\to 0)\to 0$. If we make the following estimate
\begin{equation}
\left(\frac{dS_c}{dT}\right)_{T=T_K}\simeq \frac{s_c(T_A)}{T_A-T_K}
\end{equation}
and neglect the dependence of $T_A-T_K$ on $k_0$, we find that $D\propto k_0$
in qualitative agreement with the results of Ref. \cite{Grousson2001}.

\subsection{barrier fluctuations}

Numerous experiments on supercooled liquids are not only sensitive to the
mean barrier, $\overline{F^{\ddagger }}$, but are able to measure the entire
(broad) excitation spectrum in glasses \cite{Richert02}. Most notably, the
broad peaks in the imaginary part of the dielectric function $\varepsilon
^{\prime \prime }\left( \omega \right) $ are most naturally understood in
terms of a distribution $g\left( \tau \right) $ of relaxation times, \ such
that 
\begin{equation}
\varepsilon ^{\prime \prime }\left( \omega \right) \propto \int d\tau
g\left( \tau \right) \frac{\omega \tau }{1+\left( \omega \tau \right) ^{2}}.
\end{equation}%
Similarly dynamical heterogeneity with spatially fluctuating relaxation
times yields non-exponential (frequently stretched exponential) relaxation
of the correlation function 
\begin{equation}
\phi \left( t\right) =\int d\tau g\left( \tau \right) e^{-t/\tau },
\label{phi}
\end{equation}%
Other effects that are most likely caused by a distribution of relaxation
rates include the break down of the Stokes-Einstein relation $D=\frac{k_{B}T%
}{4\pi \eta L}$ between the diffusion coefficient $D$ of a particle of size $%
L$ and the viscosity $\eta $ \cite{Fujara92,Cicerone95}. These experiments
call for a more detailed analysis of the fluctuations 
\begin{equation}
\overline{\delta F^{\ddagger 2}}\equiv \overline{F^{\ddagger 2}}-\overline{%
F^{\ddagger }}^{2}
\end{equation}%
of the activation barriers and, more generally, of the distribution function 
$p\left( F^{\ddagger }\right) $ of barriers. The latter yields the
distribution function of the relaxation times 
\begin{equation}
g\left( \tau \right) =p\left( F^{\ddagger }\right) \frac{dF^{\ddagger }}{%
d\tau }
\end{equation}%
through $\tau \left( F^{\ddagger }\right) =\tau _{0}\exp \left( \frac{%
F^{\ddagger }}{k_{B}T}\right) $. For example, in case of a Gaussian
distribution of barriers one obtains a broad, log-normal distribution of
relaxation rates:%
\begin{equation}
g\left( \tau \right) =\frac{1}{\tau \sqrt{2\pi \lambda }}\exp \left( -\frac{%
\log ^{2}\left( \tau /\overline{\tau }\right) }{2\lambda }\right) ,
\label{lognorm}
\end{equation}%
with 
\begin{equation}
\lambda =\overline{\delta F^{\ddagger 2}}/\left( k_{B}T\right) ^{2}
\end{equation}%
and $\overline{\tau }$ from Eq. (\ref{meanr}). While the distribution, Eq. (%
\ref{lognorm}), does not yield a stretched exponential form for the
correlation function, it can often be approximated by 
\begin{equation}
\phi \left( t\right) \simeq \exp \left( -(t/\overline{\tau })^{\beta
}\right) 
\end{equation}%
with $\beta =\left( 1+\lambda \right) ^{-1/2}$. Furthermore, the study of
higher order moments of $p\left( F^{\ddagger }\right) $ is important to
determine whether the distribution is indeed Gaussian or more complicated.

In Ref. \cite{Dzero09} we used the replica formalism discussed here as well
as the more elaborate replica method of formalism Refs. \cite%
{Franz95,Franz97,Barrat97} to determine higher order moments of the barrier
distribution function. Using the "thin wall" approximation for $Q({\mathbf{x}%
})$ given by Eq. (\ref{pcthinwall}) \ we obtain for the second moment 
\begin{equation}
\overline{\delta F^{\ddagger 2}}=A\left( R^{3}+\rho R^{2}\right) ,
\label{Fddag}
\end{equation}%
where $R$ is the radius of the droplet where the explicit expressions
for the coefficient $A$ and length $\rho $ are given in
Ref. \cite{Dzero09}. \ It is noteworthy that there is a surface
contribution to the moment of the barrier fluctuations that is a
consequence of correlations between droplet and homogeneous background
with overlap $Q^{\ast }$. In Ref. \cite{Dzero09} it was also shown
that the barrier distribution function $p\left( F^{\ddagger }\right) $
is Gaussian, at least if one considers the first six moments. The
skewness of the actual barrier distribution measured in structural
glasses is, in our view, an effect due to the interaction of spatially
overlapping instantons.

\bigskip

\section{Summary}

To summarize, we have discussed several models of glassy systems where the
randomness is self generated rather than induced by strong external factors.
In particular, we applied the replica formalism developed for the spin glass
systems to study the glass transition in the many-body systems at the
presence of an arbitrary weak disorder. We employed the Landau theory to
analyze the mean field glass transition using the saddle point
approximation. We have also considered the energy fluctuations around the
saddle point and evaluated the barrier height distribution.

\begin{acknowledgments}
We are grateful to Harry Westfahl Jr. for discussions and collaborations on
problems discussed in this chapter. This research was supported by the Ames
Laboratory, operated for the U.S. Department of Energy by Iowa State
University under Contract No. DE-AC02-07CH11358 (J. S.), a Fellowship of the
Institute for Complex Adaptive Matter, by the Intelligence Advanced Research
Projects Activity (IARPA) through the US Army Research Office award
W911NF-09-1-0351 and Kent State University (M.D.), and the National Science
Foundation grant CHE-0317017 (P. G. W.).
\end{acknowledgments}


\begin{thebibliography}{99}
\bibitem{Angel96} C. A. Angel, in Proceedings of the XIV Sitges Conference, 
\emph{Complex Behavior of Glassy Systems}, ed. by M. Rubi and C.
Perez-Vicente, Lecture Notes in Physics, \textbf{492}, p. 1 (1996).

\bibitem{ang88} C. A. Angell, J. Phys. Chem. Sol. \textbf{49}, 863 (1988).

\bibitem{mc} W. G\"{o}tze, in \emph{Liquids, Freezing and Glass Transition},
ed. J.-P. Hansen, D. Levesque and J. Zinn-Justin (North-Holland, Amsterdam,
1991), p. 287.

\bibitem{KT87} T. R. Kirkpatrick and D. Thirumalai, Phys. Rev. Lett. \textbf{%
58}, 2091 (1987).

\bibitem{KW87} T. R. Kirkpatrick and P. G. Wolynes, Phys. Rev. A \textbf{35}%
, 3072 (1987).

\bibitem{Sadoc} M. Kleman and J. F. Sadoc, J. Physique Lett. \textbf{40}
L569 (1979), J. F. \ Sadoc, J. Phys. Lett. 44 L707 (1983); J. F. Sadoc and
R. Mosseri, J. Physique 45 1025 (1984).

\bibitem{Nelson} D. R. Nelson, Phys. Rev. Lett. 50 982 (1983); D. R. Nelson,
Phys. Rev. B \textbf{28} 5515 (1983); S. Sachdev and D. R. Nelson, Phys.
Rev. Lett. \textbf{53} 1947 (1984); S. Sachdev and D. R. Nelson, Phys. Rev.
B \textbf{32} 1480 (1985).

\bibitem{Sethna} J. P. Sethna, Phys. Rev. Lett. \textbf{51} 2198 91983); J.
P. Sethna, Phys. Rev. B \textbf{31} 6278 (1985).

\bibitem{Sausset} F. Sausset, G. Tarjus and P. Viot, Journal of Statistical
mechanics: Theory and Experiment, P04022 (2009).

\bibitem{DKivelson} D. Kivelson, S.A. Kivelson, X. Zhao, Z. Nussinov and G.
Tarjus. Physica A \textbf{219}, 27 (1995),

\bibitem{Chayes} L. Chayes, V. J. Emery, S. A. Kivelson, Z. Nussinov and G.
Tarjus, Physica A \textbf{225} 129 (1996).

\bibitem{Nussinov} Z. Nussinov, J. Rudnick, S. A. Kivelson, and L. N.
Chayes, Phys. Rev. Lett. \textbf{83} 472 (1999).

\bibitem{Grousson} M. Grousson, G. Tarjus, and P. Viot, Phys. Rev. E \textbf{%
65}, 065103(R) (2002); M Grousson, G Tarjus and P Viot J. Phys.: Condens.
Matter \textbf{14} 1617 (2002).

\bibitem{GroussonMC} M. Grousson, V. Krakoviack, G. Tarjus, and P. Viot,
Phys. Rev. E \textbf{66}, 026126 (2002).

\bibitem{Tarjus} G. Tarjus, S. A. Kivelson, Z. Nussinov and P. Viot, J. Phys.:
Condens. Matter \textbf{17}, R1143 (2005).

\bibitem{Mon95} R. Monasson, Phys. Rev. Lett. \textbf{75}, 2847 (1995).

\bibitem{MePa98.1} M. Mezard and G. Parisi, Phys. Rev. Lett. \textbf{82},
747 (1999).

\bibitem{MePa98.2} M. Mezard and G. Parisi, J. Chem. Phys. \textbf{111},
1076 (1999).

\bibitem{Franz95} S. Franz and G. Parisi, J. Phys. I (France) \textbf{5},
1401 (1995).

\bibitem{KTW89} T. R. Kirkpatrick and D. Thirumalai, and P. G. Wolynes,
Phys. Rev. A \textbf{40}, 1045 (1989).

\bibitem{XW00} X. Xia and P. G. Wolynes, Proc. Natl. Acad. Sci. \textbf{97},
2990 (2000).

\bibitem{XW01} X. Xia and P. G. Wolynes, Phys. Rev. Lett. \textbf{86}, 5526
(2001).

\bibitem{Lubchenko01} V. Lubchenko, P. G. Wolynes, Phys. Rev. Lett. \textbf{%
87}, 195901 (2001).

\bibitem{Lubchenko04a} V. Lubchenko, P. G. Wolynes, \ Journ. of Chem. Phys. 
\textbf{121}, 2852 (2004).

\bibitem{Biroli04} G. Biroli and J.-P. Bouchaud, Journ. of Chem. Phys., 
\textbf{121}, 7347 (2004).

\bibitem{Kauzmann48} W. Kauzmann, Chemical Reviews \textbf{43},219 (1948).

\bibitem{Langer67} J. S. Langer, Ann. Phys., NY \textbf{41,} 108 (1967).

\bibitem{stripe} J.H. Cho, F.C. Chou, and D.C. Johnston, Phys. Rev. Lett. 
\textbf{70}, 222 (1993).

\bibitem{stripe2} J.M. Tranquada, J.J. Sternlieb, J.D. Axe, Y. Nakamura, and
S. Uchida, Nature (London) \textbf{375}, 561 (1995).

\bibitem{mang01} D. N. Argyriou, J. W. Lynn, R. Osborn, B. Campbell, J. F.
Mitchell, U. Ruett, H. N. Bordallo, A. Wildes, C. D. Ling, Physical Review
Letters \textbf{89}, 036401 (2002).

\bibitem{Millis96} A. J. Millis, Phys. Rev. B \textbf{53}, 8434 (1996).

\bibitem{Dagotto} E. Dagotto, T. Hotta, and A. Moreo, Physics Reports (2001).

\bibitem{nmr01} M.-H. Julien, F. Borsa, P. Carretta, M. Horvatic, C.
Berthier, and C. T. Lin, Phys. Rev. Lett. \textbf{83}, 604 (1999).

\bibitem{nmr02} A. W. Hunt, P. M. Singer, K. R. Thurber, and T. Imai, Phys.
Rev. Lett. \textbf{82}, 4300 (1999).

\bibitem{nmr03} N. J. Curro, P. C. Hammel, B. J. Suh, M. H\"{u}cker, B. B%
\"{u}chner, U. Ammerahl, and A. Revcolervschi, Phys. Rev. Lett. \textbf{85},
642 (2000).

\bibitem{nmr3b} J. Haase, R. Stern, C. T. Milling, C. P. Slichter, and D. G.
Hinks, Physica C \textbf{341}, 1727 (2000).

\bibitem{nmr04} N. J. Curro, Journal of Physics and Chemistry of Solids 
\textbf{63}, 2181(2002).

\bibitem{msr01} Ch. Niedermeyer, C. Bernhard, T. Blasius, A. Golnik, A.
Moodenbaugh, and J. I. Budnik, Phys. Rev. Lett. \textbf{80}, 3843 (1998).

\bibitem{msr02} C. Panagopoulos, J. L. Tallon, B. D. Rainford, \ T.Xiang, J.
R. Cooper, and C. A. Scott, Phys. Rev. B \textbf{66}, 064501 (2002).

\bibitem{Popovich07} J. Jaroszy\'{n}ski, T. Andrearczyk, G. Karczewski, J. Wr%
\'{o}bel, T. Wojtowicz, Dragana Popovi\'{c}, and T. Dietl, Phys. Rev. B 
\textbf{76}, 045322 (2007)

\bibitem{Spivak2006} B. Spivak, S. A. Kivelson, Annals of Physics \textbf{321%
}, 2071 (2006).

\bibitem{2DSpivak2010} B. Spivak, S. V. Kravchenko, S. A. Kivelson, X. P. A.
Gao, Rev. Mod. Phys. \textbf{82}, 1743 (2010).

\bibitem{QHE} K.B. Cooper, M.P. Lilly, J.P. Eisenstein, P.N. Pfeiffer, and
K.W. West, Phys. Rev. B \textbf{60}, 11285 (1999).

\bibitem{QHE2} S.A. Parameswaran, S.A. Kivelson, S.L. Sondhi, B.Z. Spivak, 
\emph{pre-print} arXiv:1010.4908 (2010).

\bibitem{magmtlyr} T. Garel and S. Doniach, Phys. Rev. B \textbf{26}, 325
(1982); R. Allenspach and A. Bischof, Phys. Rev. Lett. \textbf{69}, 3385
(1992).

\bibitem{selfassbl} P.G. deGennes and C. Taupin, J. Phys. Chem. \textbf{86},
2294 (1982); W.M. Gelbart and A. Ben Shaul, J. Phys. Chem. \textbf{100},
13169 (1996).

\bibitem{Ohta86} T. Ohta and K. Kawasaki, Macromolecule \textbf{19}, 2621
(1986).

\bibitem{Popovic2002} S. Bogdanovich and D. Popovic, Phys. Rev. Lett. 
\textbf{88}, 236401 (2002).

\bibitem{Popovic2} J. Jaroszynki and D. Popovic, Phys. Rev. Lett. \textbf{96}%
, 037403 (2006).

\bibitem{Vlad2002} C. Panagopoulos and V. Dobrosavljevic, Phys. Rev. \textbf{%
B72}, 014536 (2002).

\bibitem{Denis2002} D. Dalidovich and P. Philips, Phys. Rev. Lett. \textbf{89%
}, 27001 (2002).

\bibitem{Kivelson} V.J. Emery and S.A. Kivelson, Physica C \textbf{209}, 597
(1993).

\bibitem{Schmalian} J. Schmalian and P.G. Wolynes, Phys. Rev. Lett. \textbf{%
85}, 836 (2000).

\bibitem{SWW00} H. Westfahl Jr., J. Schmalian, and P. G. Wolynes, Phys. Rev.
B \textbf{64}, 174203 (2001).

\bibitem{WSW03} H. Westfahl Jr., J. Schmalian, and P. G. Wolynes, Phys. Rev.
B \textbf{68}, 134203 (2003).

\bibitem{WSW04} S.Wu, J. Schmalian, G.Kotliar, and P. G. Wolynes, Phys. Rev.
B \textbf{70}, 024207 (2004).

\bibitem{CIS98} P. Chandra, L.B. Ioffe, and D. Sherrington, Phys. Rev. B 
\textbf{58}, R14669 (1998).

\bibitem{Dzero05} M. Dzero, J. Schmalian, and P. G. Wolynes, Phys. Rev. B 
\textbf{72}, 100201 (2005).

\bibitem{Dzero09} M. Dzero, J. Schmalian, and P. G. Wolynes, Phys. Rev. B 
\textbf{80}, 024204 (2009).

\bibitem{Leibler} L. Leibler, Macromolecules \textbf{13}, 1602 (1980).

\bibitem{Fredrickson} G. H. Fredrickson and E. Helfand, J. Chem. Phys. 
\textbf{87}, 697 (1987).

\bibitem{Deem94} M. W. Deem and D. Chandler, Phys. Rev. E \textbf{49}, 4268
(1994).

\bibitem{WWSW02} S. Wu, H. Westfahl Jr., J. Schmalian, and P. G. Wolynes,
Chem. Phys. Lett. \textbf{359}, 1 (2002).

\bibitem{Zhang06} C. Z. Zhang and Z. G. Wang, Phys. Rev. E \textbf{73},
031804 (2006).

\bibitem{Wu09} S. Wu, Phys. Rev. E 79, 031803 (2009).

\bibitem{Tarzia06} M. Tarzia and A. Coniglio, Phys. Rev. Lett.\textbf{\ 96},
075702 (2006).

\bibitem{Tarzia07} M. Tarzia and A. Coniglio, Phys. Rev. E \textbf{75},
011410 (2007)

\bibitem{Brazovskii} S. A. Brazovskii, Zh. Exp. Teor. Fiz. \textbf{68}, 175
(1975)[Sov. Phys. JEPT \textbf{41}, 85 (1975)].

\bibitem{BDM87} S. A. Brazovskii, I. E.Dzyaloshinskii and A. R. Muratov,
Sov. Phys. JETP, \textbf{66} 635 (1987).

\bibitem{Alexander78} S. Alexander and J. McTague, Phys. Rev. Lett. \textbf{%
41}, 702 (1978).

\bibitem{Groh99} B. Groh and B. Mulder, Phys. Rev. E \textbf{59}, 5613
(1999).

\bibitem{Klein01} W. Klein, Phys. Rev. E \textbf{64}, 056110 (2001).

\bibitem{Geissler04} P. L. Geissler and D. R. Reichman, Phys. Rev. E \textbf{%
69}, 021501 (2004)

\bibitem{Hansen} J. P. Hansen and I. R. McDonald, \emph{Theory of simple
liquids}, Academic Press, London, N.Y., San Francisco, 1976.

\bibitem{RY79} T. V. Ramakrishnan and M. Yussouff, Phys. Rev. B \textbf{194}%
, 2775 (1979); M. Youssouff, Phys. Rev. B \textbf{23}, 5871 (1983).

\bibitem{Stoessel84} J. P. Stoessel and P. G. Wolynes, J. Chem. Phys. 
\textbf{80}, 4502 (1984).

\bibitem{Singh85} Y. Singh, J. P. Stoessel, P. G. Wolynes, Phys. Rev. Lett. 
\textbf{54}, 1059 (1985).

\bibitem{N98} Th. M. Nieuwenhuizen, Phys. Rev. Lett. \textbf{80}, 5580
(1998).

\bibitem{Landau} L. D. Landau and E. M. Lifshitz, \emph{Statistical Physics}
(Pergamon London 1980).

\bibitem{Donth} E. Donth, J. Non-Cryst. Solids \textbf{53} 325 (1982).

\bibitem{CK93} L. F. Cugliandolo, J. Kurchan, Phys. Rev. Lett. \textbf{71},
173 (1993).

\bibitem{SCSA} A.J. Bray, J. Phys. A: Math., Nucl. Gen. \textbf{7}, 2144
(1974).

\bibitem{Gross85} D. J. Gross, I. Kanter, and H. Sompolinsky, Phys. Rev.
Lett. \textbf{55}, 304 (1985).

\bibitem{Franz05-1} S. Franz, J. Stat. Mech.-Theory and Exp. p04001 (2005).

\bibitem{Franz97} S. Franz and G. Parisi, Phys. Rev. Lett. \textbf{79}, 2486
(1997).

\bibitem{Barrat97} A. Barrat, S. Franz, and G. Parisi, J. Phys. A: Math.
Gen. \textbf{30}, 5593 (1997).

\bibitem{Richert02} R. Richert, J. Phys. Cond. Matt. \textbf{24} R703 (2002).

\bibitem{Fujara92} F. Fujara, B. Geil, H. Sillescu and G. Fleischer,
Zeitschr. f. Phys. B \textbf{88}, 195 (1992).

\bibitem{Cicerone95} M. T. Cicerone, M. D. Ediger, Journ. of Chem. Phys., 
\textbf{104}, 7210 (1996).

\bibitem{Grousson2001} M. Grousson, G. Tarjus, and P. Viot, Phys. Rev. Lett. 
\textbf{86}, 3455 (2001).
\end{thebibliography}
\end{document}